\documentclass[english,12pt]{article}
\usepackage[T1]{fontenc}
\usepackage[latin9]{inputenc}
\usepackage{nicefrac}
\usepackage{geometry}
\usepackage{float}
\usepackage{bm}
\usepackage{amsmath}
\usepackage{afterpage}
\usepackage{lscape}
\usepackage{amssymb}
\usepackage{graphicx}
\usepackage[authoryear]{natbib}
\usepackage{listings}
\usepackage{courier}
\usepackage{algorithm}
\usepackage{algorithmicx}
\usepackage{hhline}
\usepackage{colortbl}
\usepackage{xcolor}
\usepackage{multirow}
\usepackage{cancel}
\usepackage[section]{placeins}

\usepackage{babel}
\pdfoutput=1
\begin{document}

\def\spacingset#1{\renewcommand{\baselinestretch}%
{#1}\small\normalsize} \spacingset{1}

\newcommand{\blind}{0}


\if0\blind
{
  \title{\bf Detection of latent heteroscedasticity and group-based regression effects in linear models via Bayesian model selection}
  \author{Thomas A. Metzger\hspace{.2cm}\\
    Department of Statistics, Virginia Tech\\
    and \\
    Christopher T. Franck\\
    Department of Statistics, Virginia Tech}
  \maketitle
} \fi

\if1\blind
{
  \bigskip
  \bigskip
  \bigskip
  \begin{center}
    {\LARGE\bf Detection of latent heteroscedasticity and group-based regression effects in linear models via Bayesian model selection}
\end{center}
  \medskip
} \fi

\bigskip
\begin{abstract}
Standard linear modeling approaches make potentially simplistic assumptions regarding the structure of categorical effects that may obfuscate more complex relationships governing data. For example, recent work focused on the two-way unreplicated layout has shown that hidden groupings among the levels of one categorical predictor frequently interact with the ungrouped factor. We extend the notion of a ``latent grouping factor'' to linear models in general. The proposed work allows researchers to determine whether an apparent grouping of the levels of a categorical predictor reveals a plausible hidden structure given the observed data. Specifically, we offer Bayesian model selection-based approaches to reveal latent group-based heteroscedasticity, regression effects, and/or interactions. Failure to account for such structures can produce misleading conclusions. Since the presence of latent group structures is frequently unknown \textit{a priori} to the researcher, we use fractional Bayes factor methods and mixture $g$-priors to overcome lack of prior information. 
\end{abstract}

\noindent%
\vfill
\newpage
\spacingset{2} 

\section{Introduction}\label{sec:intro}
\begin{figure}[htp] 
\begin{centering}
\includegraphics{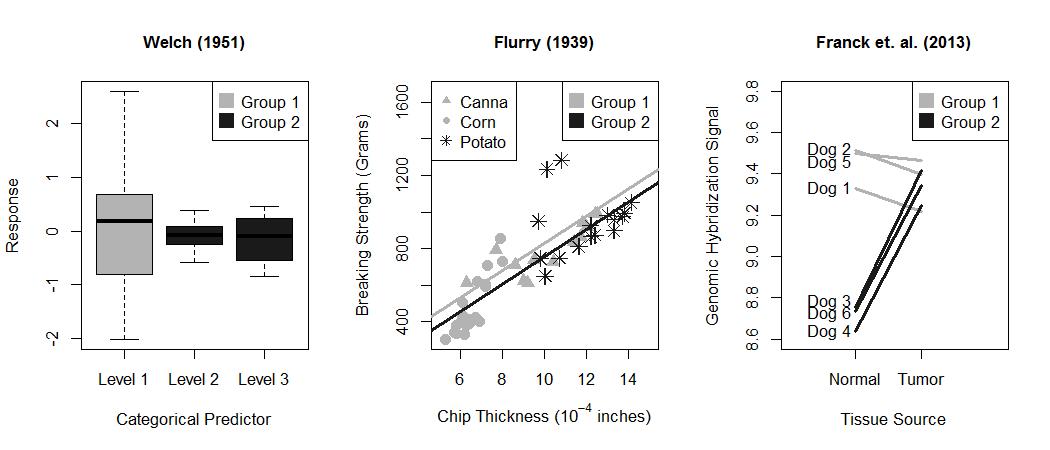}
\par\end{centering}
\caption{Data from \citet{Welch1951Heteroanova} (left), \citet{Flurry} (center), and \citet{Franck2013} (right). After a cursory examination of the data, a researcher might suspect that a latent grouping factor (emphasized by dark and light gray) underlies the levels of the categorical predictor.}
\label{fig:threeplots}
\end{figure}

Linear models with categorical predictors are among the most frequently used statistical models, but oversimplification of the variance or regression effect structures can misrepresent key relationships within observed data. Figure \ref{fig:threeplots} shows three relevant data sets. First, the left panel shows a simulated one-way ANOVA experiment, based on the statistics reported in \citet{Welch1951Heteroanova}. The horizontal axis represents the levels of a treatment factor, and the vertical axis represents a continuous response. The center panel is an analysis of covariance (ANCOVA) that analyzes the breaking strength of a starch chip \citep{Flurry}. The horizontal axis represents the chip's thickness in $10^{-4}$ inches, the vertical axis represents the breaking strength in grams, and the point shapes represent the plant from which the starch was derived. Finally, the right panel is an unreplicated two-way layout, measuring the genomic hybridization signal in dogs with lymphoma \citep{Franck2013}. The horizontal axis represents whether the sample was taken from normal or tumor tissue, the vertical axis represents the intensity of the genomic hybridization signal, and the individual lines represent six dogs studied. In each case, the levels of the categorical predictor appear to fall into one of two groups, although the group structure is unknown before data collection. The apparent group structure for the data in Figure \ref{fig:threeplots} is represented by dark and light gray. 

Situations similar to those illustrated in Figure \ref{fig:threeplots} often arise in research. Perhaps after plotting the data or in reviewing previous related work, the researcher begins to suspect that there is a hidden grouping within the levels of the categorical predictor. We call this predictor the \textit{suspected latent grouping factor}, or SLGF. This work aims to determine whether the latent groupings are plausible and how the group structure affects the response. 

We must consider the SLGF with two key ideas in mind: first, that the SLGF might manifest itself through one of several \textit{structures} in the data, namely, (i) group-specific regression effects, (ii) group-specific variances, and/or (iii) hidden interactions between groups and other model predictors. There are eight possible combinations of structures. Second, the specific assignment of levels to groups is unknown. Both of these aspects of the SLGF must be learned from the data in an unsupervised fashion: specifically, our proposed method uses Bayesian model selection to assess the plausibility of the SLGF's impact on the data based on posterior probability. Formal detail can be found in Section 2. 


Common linear model assumptions include homoscedasticity and a unique effect of the response at each level of categorical predictors; for a thorough review see \citet{Berry}. Violations of these assumptions can have a wide variety of negative consequences on inference; see \citet{RencherandSchaaljeLMs}, \citet{Deschamps1991} and \citet{ScheffeANOVA}. Model misspecification can lead to additional problems; see \citet{RencherandSchaaljeLMs}, \citet{PRao1971Misspecification}, and \citet{Deegan}. 

Regarding the detection of heteroscedasticity, many analyses follow a two-stage approach. These include the methods proposed by \citet{Bartlett1937}, \citet{Levene1960}, \citet{BrownForsythe}, and \citet{Hartley}. When heteroscedasticity is believed to be a function of a continuous predictor, many methods are available, including \citet{BreuschPagan}, \citet{CookWeisberg}, \citet{White1980}, \citet{Glejser1969}, \citet{Park}, \citet{BoxandHill}, \citet{Bickel}, \citet{Jobson}, and \citet{CarrollRuppert1982}.

Regarding inference in the presence of heteroscedasticity, methods include those of \citet{BoxCoxTransform}, \citet{CR1988}, \citet{GaussWLS,MorrisonLMsText}, \citet{huber1967}, \citet{eicker1967}, \citet{Cragg}, \citet{HildrethHouck1968}, \citet{LongErvin}, \citet{Polasek2}, \citet{Polasek}, \citet{Cuervo}, \citet{BoscardinGelman}, \citet{mackinnonwhite}, \citet{Dumitrascu}, and \citet{White1980}; for a review see \citet{HayesOSU} and \citet{NetoBoot}. Many methods exist to detect heteroscedasticity or conduct inference in its presence. To our knowledge, ours is the first proposal of latent group-based heteroscedasticity alongside possibly unique regression effects and/or hidden interactions. 

While an extension to more than two groups is natural, our choice of two groups is still a reasonable approach in many problems; \citet{KKSA} and \citet{Franck2018} study factor level groupings based on two groups in unreplicated two-way layouts, while \citet{GoldfeldQuandt} model heteroscedasticity as a function of two groups, where groups are created by partitioning observations ordinally. 

The main contribution of this work is to propose a method of probabilistically detecting the presence of hidden categorical level groupings, and to describe the model specifications that capture the effect of such groupings, through the use of Bayesian model selection. This work generalizes the notion of latent group-based effects from the two-way layout to linear models; see \citet{KKSA, Franck2013, FranckRJ}, and \citet{Franck2018}. There is no unified Bayesian model selection approach to account for these structures in the context of latent groupings of the levels of a categorical predictor in general.  Although we illustrate our method in the contexts of the three specific settings shown in Figure \ref{fig:threeplots}, our proposal is flexible enough to be used in the context of any linear model with a categorical predictor. 

The remaining structure of this paper is as follows. Section 2 describes the candidate models in the context of the SLGF, as well as the fractional Bayes factor and Bayesian model selection details. Section 3 describes our proposed Bayesian model specification in contexts of ANCOVA models and unreplicated two-way layouts. 
Section 4 describes a simulation study to assess the performance of our method. Section 5 applies our proposed method to the empirical data sets of Figure \ref{fig:threeplots}. Section 6 summarizes the proposed method and provides some additional comments. Additional mathematical details and simulation results for one-way ANOVA data are provided in the supplement. 

\section{Proposed Method}
\subsection{Specification of Linear Models with Categorical Predictors}\label{sec:linearmodelspec}
We begin the development of our approach by elucidating the assignment of the levels of the SLGF into groups. As an illustration, consider the data analyzed by \citet{Franck2013}, shown in the right panel of Figure  \ref{fig:threeplots}: an unreplicated two-way layout with 6 rows and 2 columns. We choose the dogs as the SLGF. Three examples of possible SLGF level assignments are shown in Figure \ref{fig:groupingexamples}: 

\begin{figure}[htp] 
\begin{centering}
\begin{tabular}{|c||c|c|}
\hline 
\rowcolor{gray}$r_{1}$ & 9.33 & 9.22\tabularnewline
\hline 
\rowcolor{gray}$r_{2}$ & 9.51 & 9.39\tabularnewline
\hline 
\rowcolor{gray}$r_{3}$ & 8.75 & 9.42\tabularnewline
\hline 
$r_{4}$ & 8.64 & 9.25\tabularnewline
\hline 
$r_{5}$ & 9.50 & 9.46\tabularnewline
\hline 
$r_{6}$ & 8.73 & 9.35\tabularnewline
\hline 
\multicolumn{3}{|c|}{Example scheme 1}\tabularnewline
\hline 
\end{tabular}\qquad{}%
\begin{tabular}{|c||c|c|}
\hline 
\rowcolor{gray}$r_{1}$ & 9.33 & 9.22\tabularnewline
\hline 
\rowcolor{gray}$r_{2}$ & 9.51 & 9.40\tabularnewline
\hline 
$r_{3}$ & 8.75 & 9.42\tabularnewline
\hline 
\rowcolor{gray}$r_{4}$ & 8.64 & 9.25\tabularnewline
\hline 
$r_{5}$ & 9.50 & 9.46\tabularnewline
\hline 
\rowcolor{gray}$r_{6}$ & 8.73 & 9.34\tabularnewline
\hline 
\multicolumn{3}{|c|}{Example scheme 2}\tabularnewline
\hline 
\end{tabular}\qquad{}%
\begin{tabular}{|c||c|c|}
\hline 
\rowcolor{gray}$r_{1}$ & 9.33 & 9.22\tabularnewline
\hline 
$r_{2}$ & 9.51 & 9.39\tabularnewline
\hline 
$r_{3}$ & 8.75 & 9.42\tabularnewline
\hline 
\rowcolor{gray}$r_{4}$ & 8.64 & 9.25\tabularnewline
\hline 
\rowcolor{gray}$r_{5}$ & 9.50 & 9.46\tabularnewline
\hline 
$r_{6}$ & 8.73 & 9.34\tabularnewline
\hline 
\multicolumn{3}{|c|}{Example scheme 3}\tabularnewline
\hline 
\end{tabular}
\par\end{centering}
\caption{Three possible grouping schemes of the data analyzed by \citet{Franck2013} (as shown in the rightmost panel of Figure  \ref{fig:threeplots}) are shown here. Row membership is used to partition the data into two groups, shaded and unshaded. These grouping schemes are denoted $1,2,3:4,5,6$ (left), $1,2,4,6:3,5$ (center), and $1,4,5:2,3,6$ (right).}
\label{fig:groupingexamples}
\end{figure}
For a SLGF with levels $k=1,\ldots ,K+1$, let $\bm{k}=(1,\dots,K+1)^T$. We formally define a \textit{grouping scheme}, denoted ${d_s}(\bm{k})$, as a partitioning that assigns the data into two groups based on each observation's corresponding level of the SLGF. Let $\mathcal{S}$ be the set of all possible schemes, and $s=1,\dots,S$ index the possible schemes. We refer to the example scheme 3, in the rightmost panel of Figure \ref{fig:groupingexamples}, to motivate the subsequent notational definitions; this example partitions levels $k=\ $1, 4, and 5 separately from 2, 3, and 6. For a scheme ${d_s}(\bm{k})$, a colon separates the levels each group comprises; for example, the scheme of example 3 is denoted as ${d_s}(\bm{k})=1,4,5:2,3,6$. The most effective method to partition the levels of the SLGF into groups depends on the size and nature of the study in question. Many problems have $K$ small enough that a combinatoric search over all possible grouping schemes is reasonable; see \citet{Franck2013}, \citet{FranckRJ}, \citet{Franck2018}, and \citet{KKSA}. We use the combinatoric search approach exclusively in this study. 

Next we formalize the idea of model structures in the context of  specific linear models. Recall from Section \ref{sec:intro} that we must potentially accommodate eight model structures containing a mix of group-based regression effects, group-based variances, and group-based interactions. These structures must be tailored to both the researcher's suspicion and goals, as well as the data layout under consideration. Thus we propose the \textit{model class}, which is the set of models representing a particular structure. While structures reflect the presence or absence of group-based effects, classes prescribe specific corresponding models in the context of the data layout in question. 

Now that the grouping schemes and model classes have been enumerated, consider a linear model with $N$ centered observations $\boldsymbol{Y}$:
\begin{equation}
\boldsymbol{Y}=X\boldsymbol{\beta}+\bm{\varepsilon}\label{eq:Wmodel}
\end{equation}
with model matrix $X$ parametrized to be full column rank, regression effects $\boldsymbol{\beta}$,
and errors $\boldsymbol{\varepsilon}\overset{\text{iid}}{\sim}N(\boldsymbol{0},\,\Sigma)$
with covariance matrix $\Sigma$. To account for the eight possible model structures previously described, we will partition $\bm{Y},\, \boldsymbol{\beta},\, \boldsymbol{\varepsilon},$ and $\Sigma$. Partition $\boldsymbol{\beta}$ into four components: let $\alpha$ represent an intercept common to all models, let $\boldsymbol{\nu}=\{\nu_{k}\}_{k=1}^{K}$ represent the SLGF with $K+1$ levels, let $\boldsymbol{\tau}=\{\tau_{j}\}_{j=1}^{J}$ contain $J$ other regression effects, categorical or continuous, and, let $\boldsymbol{\rho}=\{\rho_\ell\}_{\ell=1}^L$ represent $L$ interactions with the SLGF. Then $\boldsymbol{\beta}_{(1+K+J+L)\times 1}=(\alpha,\ \boldsymbol{\nu}_{K\times 1},\ \boldsymbol{\tau}_{J\times 1},\, \boldsymbol{\rho}_{L\times 1})$. Similarly, partition the model matrix $X_{N\times (1+K+J+L)}=(\bm{1}^T_{N\times 1}\,|W_{N\times K}\,|\,V_{N\times J}\,|\,U_{N\times L})$ into three matrices corresponding to the data related to $\alpha,\, \boldsymbol{\nu}$, $\boldsymbol{\tau}$, and $\boldsymbol{\rho}$, respectively. 
Thus we can express (\ref{eq:Wmodel}) equivalently as 
\begin{equation}
\boldsymbol{Y}=\bm{1}^T \alpha + W\boldsymbol{\nu}+V\boldsymbol{\tau}+U\boldsymbol{\rho}+\boldsymbol{\varepsilon}.\label{eq:WVUmodel}
\end{equation}

In cases where the effect structure for one of the terms in Equation  (\ref{eq:WVUmodel}) depends on a latent grouping scheme, denote that term with a tilde. For example, in structures with group-based interactions, model a group-based interaction $\tilde{\boldsymbol{\rho}}$ instead of usual the interaction $\boldsymbol{\rho}$. Similarly, heteroscedastic structures with group-based variances are modeled with error vector $\tilde{\boldsymbol{\varepsilon}}$ and corresponding covariance matrix $\tilde{\Sigma}$ instead of the homoscedastic counterparts $\boldsymbol{\varepsilon}$ and $\Sigma$. 



When a model contains a group structure, we arrange the observations within $\bm{Y}$ to first contain the $n_1$ observations corresponding to $s_1$, followed by the $n_2$ observations corresponding to $s_2$, denoted $\bm{Y}=(\bm{Y}_1,\bm{Y}_2)$. We similarly arrange the rows and columns of $X$ to contain the corresponding observations, which helps concisely express our proposed forms of heteroscedasticity in $\Sigma$. For structures with distinct regression effects by group, arrange $\tilde{\boldsymbol{\nu}}=(\tilde{\boldsymbol{\nu}}_{1},\,\tilde{\boldsymbol{\nu}}_{2})$, and for classes with distinct variances by group, partition the error vector and covariance matrix such that $\boldsymbol{\tilde{\varepsilon}}\sim N(\bm{0},\tilde{\Sigma})$ where $\ensuremath{\tilde{\Sigma}=\left(\begin{array}{c|c}
\sigma^2_1I_{n_{1}\times n_{1}} & 0_{n_{1}\times n_{2}}\\
\hline 0_{n_{2}\times n_{1}} & \sigma^2_2I_{n_{2}\times n_{2}}
\end{array}\right)_{N\times N}}$. Notice $N=n_{1}+n_{2}$, for the effects corresponding to groups $s_{1}$ and $s_{2}$, respectively. 




With multiple schemes, classes, and structures under consideration, we next propose a Bayesian model selection approach to assess whether latent groups exist within the data, and if so, identify the appropriate grouping scheme ${d_s}(\bm{k})\in\mathcal{S}$ (if present) and class $c\in\mathcal{C}$. 



\subsection{Bayesian Model Selection Details}\label{sec:modelselection}
\subsubsection{Model Specification}\label{sec:modelspec}
Denote the set of all candidate models $\mathcal{M}=\{m_s^c\}$, indexed over all possible schemes $s=1,...,{S}$ and classes $c=1,...,{C}$, where $|\mathcal{M}|=M$; to ease the notational burden, we have denoted ${d_s}(\bm{k})$ as $s$ in the subscript of $m$. Although each model matrix $X$ and estimators for $\boldsymbol{\beta}$ and $\Sigma$ depend on the grouping and class under consideration, for notational simplicity we do not index $X$, $\boldsymbol{\beta}$, or $\Sigma$ by $m$, $s$, or $c$. Let the vector $\boldsymbol{\varphi}$ contain the single precision $\varphi:=\frac{1}{\sigma^2}$ under homoscedastic models, and the corresponding subgroup precisions $\varphi_1$ and $\varphi_2$ under heteroscedastic models. Denote the precision matrix $\Phi=\Sigma^{-1}$ and $\bm{\theta}=\{\boldsymbol{\varphi},\boldsymbol{\beta}\}$, the set of unknown model parameters; then we can express (\ref{eq:Wmodel}) conditionally as 
\begin{equation}
\bm{Y}|m_{s}^{c},\bm{\theta}\sim N(X\boldsymbol{\beta},\Phi^{-1})
\end{equation}
for a given scheme ${d_s}(\bm{k})\in \mathcal{S}$ and class $c\in \mathcal{C}$. Thus the likelihood function is given by
\begin{equation}
P(\bm{Y}|\bm{\theta},m_{s}^{c})=(2\pi)^{-\frac{N}{2}}|\Phi|^{\frac{1}{2}}\cdot\exp\{-\frac{1}{2}(\bm{Y}-X\boldsymbol{\beta})^{T}\Phi(\bm{Y}-X\boldsymbol{\beta})\}
\end{equation}
We consider two common prior specifications on the regression effects and precision(s). In both cases, we prefer noninformative priors on the precision(s) because prior information on precision is rarely available. First, we consider a noninformative approach, where we have
\begin{equation}
P(\boldsymbol{\boldsymbol{\beta},\boldsymbol{\varphi},|m_s^c})\propto {\varphi}^{-1} \text{ (homoscedastic models), or}
\label{eq:prior1-1}
\end{equation}
\begin{equation}
P(\boldsymbol{\boldsymbol{\beta},\boldsymbol{\varphi},|m_s^c})\propto {\varphi_1}^{-1}\cdot {\varphi_2}^{-1} \text{ (heteroscedastic models)}
\label{eq:prior1-2}
\end{equation}
Next we consider the Zellner-Siow mixture $g$-prior \citep{Zellner1980, Zellner1986, Liangetal}, where
\begin{equation}
P(\alpha,\,\boldsymbol{\varphi})\propto \varphi^{-1} \text{ (homoscedastic models),}
\label{eq:prior2-1}
\end{equation} 
\begin{equation}
P(\alpha,\,\boldsymbol{\varphi})\propto \varphi_1^{-1}\cdot \varphi_2^{-1} \text{ (heteroscedastic models),}
\label{eq:prior2-2}
\end{equation} 
\begin{equation}
\boldsymbol{\beta}_{- \alpha}|\boldsymbol{\varphi},m_s^c\sim N(\bm{0},\ g(X^T\Phi^{-1} X)^{-1})\text{, and}
\label{eq:prior2-3}
\end{equation} 
\begin{equation}
g\sim \text{IG} \left( \frac{1}{2},\ \frac{N}{2}\right)
\label{eq:prior2=4}
\end{equation}
where $\boldsymbol{\beta}_{-\alpha}:=\boldsymbol{\beta}\setminus \{\alpha\}$. We use the Zellner-Siow mixture $g$-prior in relatively data-poor situations, such as unreplicated two-way layouts, to reduce the dimensionality of the parameter space with improper priors. This consequently lowers the minimal training sample size, a critical component of the fractional Bayes factor approach described in Section \ref{sec:fbfs}. We use the noninformative flat prior where data are more abundant; see Section \ref{sec:twowayms} for more detail.

\subsubsection{Model Priors for Classes and Schemes}
We impose a uniform model prior by model class: $P(m^{c})=\sum_{s=1}^{S}P(m_{s}^{c}):=\frac{1}{C}$. Depending on the data layout and model structures under consideration, various classes may contain different numbers of models; thus we subsequently divide each class's prior uniformly among the models it contains. For example, in an ANOVA layout, one model class might represent the single mean model, where $P(m^c)=\frac{1}{C}$; note we do not index this model by $s$ as there is no grouping structure present in this class. Alternatively, for a model class containing models with distinct regression effects by grouping scheme, the $S$ individual models within the class would be given the prior $P(m_{s}^c)=\frac{1}{S\cdot C}$. By implementing Bayes' Theorem, posterior model probabilities approximated via the fractional Bayes factor (see Section \ref{sec:fbfs}) can then be easily computed for each individual model and class with the marginal probabilities of each model. For a given model $m_{c^{\prime}}^{s^{\prime}}$ we compute: 
\begin{equation} 
P(m_{s^{\prime}}^{c^{\prime}}|\boldsymbol{Y})=\frac{P(\bm{Y}|m_{s^{\prime}}^{c^{\prime}})P(m_{s^{\prime}}^{c^{\prime}})}{\overset{C}{\underset{c=\text{I}}{\sum}}\overset{S}{\underset{s=1}{\sum}}P(\bm{Y}|m_{s}^{c})P(m_{s}^{c})}
\end{equation}
We can then aggregate the overall probability of a given class $c^{\prime}$ as $P(m^{c^{\prime}}|\bm{Y})=\underset{s\in\mathcal{S}}{\sum}P(m_{s}^{c^{\prime}}|\bm{Y})$, or for a given grouping scheme $s^{\prime}$ as $P(m_{s^{\prime}}|\bm{Y})=\underset{c\in\mathcal{C}}{\sum}P(m_{s^{\prime}}^{c}|\bm{Y})$; hence the proposed method allows researchers to draw probabilistic conclusions about both structures and specific grouping schemes. 
\subsubsection{Fractional Bayes Factor Approach}\label{sec:fbfs}
We next motivate the need to use a fractional Bayes factor approach. Consider a comparison between arbitrary models $m_1$ and $m_2$, where $m_1$ and $m_2$ represent homoscedastic and heteroscedastic models, respectively, via $B^{12}=\frac{P(m_1|\boldsymbol{Y})P(m_{2})}{P(m_2|\boldsymbol{Y})P(m_{1})}$. For $m_1$ we use $P(\boldsymbol{\varphi})= a\cdot \varphi^{-1}$, and for $m_2$ we use $P(\boldsymbol{\varphi})= a^{\prime} \cdot \varphi_1^{-1}\cdot \varphi_2^{-1}$ for arbitrary constants $a\ne a^{\prime}$. So the Bayes factor to compare $m_1$ and $m_2$ is  
\begin{equation}
B^{12}=\frac{a \cdot\int P(\bm{Y}|\boldsymbol{\beta},\Phi,m_1)P(\boldsymbol{\beta})P(\boldsymbol{\varphi})d\boldsymbol{\beta} d\boldsymbol{\varphi}\cdot P(m_2)}{a^{\prime}\cdot\int P(\bm{Y}|\boldsymbol{\beta},\Phi,m_2)P(\boldsymbol{\beta})P(\boldsymbol{\varphi})d\boldsymbol{\beta} d\boldsymbol{\varphi}\cdot P(m_1)}
\end{equation}
which is defined only up to the arbitrary constant $\frac{a}{a^{\prime}}$ and thus inappropriate for use in model comparison. Note this problem arises from the use of noninformative priors on the precisions when comparing homoscedastic and heteroscedastic models. Fractional Bayes factors were developed to elicit a cancellation of this constant, rendering a Bayes factor that is well-defined \citep{OHaganfbfs}. 

For a description of fractional Bayes factors (FBFs), see \citet{OHaganfbfs} and \citet{OHagan1997}. We use the FBF approach of O'Hagan rather than the intrinsic Bayes factor of \citet{IntrinsicBFsBergerandPericchi}, as the need to choose a training sample would be complicated by the potential scarcity of data induced by some clustering schemes. 

To fully quantify the fractional marginal probability of a given model $m_i$, we must compute both \linebreak $\int P(\bm{Y}|\boldsymbol{\theta}_i,m_i)\pi(\boldsymbol{\theta}_i)d\boldsymbol{\theta}_i$ and $\int P^b(\bm{Y}|\boldsymbol{\theta}_i,m_i)\pi(\boldsymbol{\theta}_i)d\boldsymbol{\theta}_i$ for some user-chosen fractional exponent $b$. \citet{OHaganfbfs} provides several recommendations for \textbf{$b$}, including $b:=\frac{m_{0}}{N}$, where $m_{0}$ is the minimal training sample size necessary for $P^{b}(\bm{Y}| m_i)$ to be proper. In this study we choose $b=\frac{m_{0}}{N}$, which elicits consistent model selection \citep{OHaganfbfs}. In many cases, tractable expressions exist for both $\int P(\bm{Y}|\boldsymbol{\theta}_i,m_i)\pi(\boldsymbol{\theta})d\boldsymbol{\theta}$ and $\int P^b(\bm{Y}|\boldsymbol{\theta}_i,m_i)\pi(\boldsymbol{\theta})d\boldsymbol{\theta}$. When these integrals are intractable, we use the Laplace approximation in the computation of both $\int P(\bm{Y}|\boldsymbol{\theta}_i,m_i)\pi(\boldsymbol{\theta})d\boldsymbol{\theta}$ and $\int P^b(\bm{Y}|\boldsymbol{\theta}_i,m_i)\pi(\boldsymbol{\theta})d\boldsymbol{\theta}$, as cubature-based approximation is computationally expensive and our study of the integrand surface indicates a suitable shape. See online Supplement for justification and further detail. 

\section{Applications}
\subsection{Application 1: ANCOVA Models}\label{sec:ancovams}
We first consider an ANCOVA scenario with continuous effect $\boldsymbol{\tau}$ and where the SLGF with $K+1$ levels corresponds to the single categorical predictor effect $\boldsymbol{\nu}$. We let $V_{N\times1}$ contain the observed continuous covariate and $W_{N\times K}$ be the appropriate categorical effect design matrix. We consider models with and without the interaction effect $\boldsymbol{\rho}$, which governs whether the linear trends share a common slope. Thus we begin with the model given by
\begin{equation}
\bm{Y} = \bm{1}^T \alpha + W\boldsymbol{\nu}+V\boldsymbol{\tau}+U\boldsymbol{\rho}+\boldsymbol{\varepsilon}
\end{equation}
yielding the likelihood function
\begin{equation}
P(\bm{Y}|m_s^c,\alpha, \bm{\nu},\bm{\tau},\boldsymbol{\rho},\boldsymbol{\varphi})=(2\pi)^{-\frac{N}{2}}|\Phi|^{\frac{1}{2}}\cdot\exp\lbrace-\frac{1}{2}(\bm{Y}-\bm{1}^T \alpha -W\boldsymbol{\nu}-V\boldsymbol{\tau}-U\boldsymbol{\rho})^T\Phi(\bm{Y}-\bm{1}^T \alpha -W\boldsymbol{\nu}-V\boldsymbol{\tau}-U\boldsymbol{\rho})\rbrace
\end{equation}
To consider the model without an interaction, let $\boldsymbol{\rho}:=\bm{0}$. For models with a group effect, let $\boldsymbol{\nu}:=\tilde{\boldsymbol{\nu}}$; for models with a group-by-continuous predictor interaction, let $\boldsymbol{\rho}:=\tilde{\boldsymbol{\rho}}$; and finally for heteroscedastic models, let $\boldsymbol{\varepsilon}:=\tilde{\boldsymbol{\varepsilon}}$. As an illustration, we consider eight distinct model classes. 
\begin{enumerate}
\item Class I ($m^{\text{I}}$): the "null" model, with no categorical or continuous covariate effects and homoscedastic error variance, contains 1 model with no grouping schemes; 

$\bm{Y}=\bm{1}^T \alpha + \boldsymbol{\varepsilon}, \boldsymbol{\varepsilon}\sim N(\bm{0},\,\sigma^{2}I)$

\item Class II ($m^{\text{II}}$): the "simple linear regression (SLR)" model, with a continuous covariate effect only and homoscedastic error variance, contains 1 model with no grouping schemes; 

$\bm{Y}=\bm{1}^T \alpha + V\boldsymbol{\tau}+\boldsymbol{\varepsilon}, \boldsymbol{\varepsilon}\sim N(\bm{0},\,\sigma^{2}I)$

\item Class III ($m^{\text{III}}$): the "ANCOVA" model with categorical and continuous covariate effects and  homoscedastic error variance, contains 1 model with no grouping schemes; 

$\bm{Y}=\bm{1}^T \alpha + W\boldsymbol{\nu}+V\boldsymbol{\tau}+\boldsymbol{\varepsilon}, \boldsymbol{\varepsilon}\sim N(\bm{0},\,\sigma^{2}I)$

\item Class IV ($m_{s}^{\text{IV}}$): the "group-contracted ANCOVA" model with group and continuous covariate effects and homoscedastic error variance, contains $2^{K-1}-1$ schemes; 

$\bm{Y}=\bm{1}^T \alpha + W\tilde{\boldsymbol{\nu}}+V\boldsymbol{\tau}+\boldsymbol{\varepsilon}, \boldsymbol{\varepsilon}\sim N(\bm{0},\,\sigma^{2}I)$

\item Class V ($m^{\text{V}}$): the "interaction ANCOVA" model with categorical and continuous covariate effects, level-based interaction, and homoscedastic error variance, contains 1 model with no grouping scheme; $\bm{Y}=\bm{1}^T \alpha + W\boldsymbol{\nu}+V\boldsymbol{\tau}+U\boldsymbol{\rho}+ \boldsymbol{\varepsilon},\  \boldsymbol{\varepsilon}\sim N(\bm{0},\,\sigma^{2}I)$

\item Class VI ($m_{s}^{\text{VI}}$): the "group-interaction" model, with group and continuous covariate effects, group-based interaction, and homoscedastic error variance, contains $2^{K-1}-1$ schemes; 

$\bm{Y}=\bm{1}^T \alpha + W\tilde{\boldsymbol{\nu}}+V\boldsymbol{\tau}+U\tilde{\boldsymbol{\rho}}+\boldsymbol{\varepsilon},\  \boldsymbol{\varepsilon}\sim N(\bm{0},\,\sigma^{2}I)$

\item Class VII ($m_{s}^{\text{VII}}$): the "heteroscedastic group-contracted" model (heteroscedastic Class IV), with group and continuous covariate effects and heteroscedastic error variance, contains $2^{K-1}-1$ schemes;

$\bm{Y}=\bm{1}^T \alpha + W\tilde{\boldsymbol{\nu}}+V\boldsymbol{\tau}+\boldsymbol{\tilde{\varepsilon}}, \  \boldsymbol{\tilde{\varepsilon}}\sim N\left(\bm{0},\,\ensuremath{\ensuremath{\tilde{\Sigma}=\left[\begin{array}{c|c}
\sigma_{1}^{2}I_{n_{1}\times n_{1}} & 0_{n_{1}\times n_{2}}\\
\hline 0_{n_{2}\times n_{1}} & \sigma_{2}^{2}I_{n_{2}\times n_{2}}
\end{array}\right]_{N\times N}}}\right)$

\item Class VIII ($m_{s}^{\text{VIII}}$): the "heteroscedastic group-interaction" model (heteroscedastic Class VI), with group and continuous covariate effects, group-based interaction, and heteroscedastic error variance, contains $2^{K-1}-1$ schemes;

$\bm{Y}=\bm{1}^T \alpha + W\tilde{\boldsymbol{\nu}}+V\boldsymbol{\tau}+U\tilde{\boldsymbol{\rho}}+\tilde{\boldsymbol{\varepsilon}},\  \boldsymbol{\tilde{\varepsilon}}\sim N\left(\bm{0},\,\ensuremath{\ensuremath{\tilde{\Sigma}=\left[\begin{array}{c|c}
\sigma_{1}^{2}I_{n_{1}\times n_{1}} & 0_{n_{1}\times n_{2}}\\
\hline 0_{n_{2}\times n_{1}} & \sigma_{2}^{2}I_{n_{2}\times n_{2}}
\end{array}\right]_{N\times N}}}\right)$

\end{enumerate}
Thus the structure of no group-based effects is represented by Classes I, II, III, and V; group-based regression effects are represented by Classes IV, VI, VII, and VIII; group-based variances are represented by Classes VII and VIII. 

We use the FBF approach outlined in Section \ref{sec:fbfs}. Typically, the most complex model considered in an ANCOVA study will have a small minimal training sample size relative to the overall sample size. Thus we choose noninformative priors on the regression effects and precision(s) as described in Equations (\ref{eq:prior1-1}) and (\ref{eq:prior1-2}), so the marginal density of the data conditional on the model is
\begin{equation}
P(\bm{Y}|m_s^c)=\iint P(\bm{Y}|\boldsymbol{\beta},\boldsymbol{\varphi},m_s^c)P(\boldsymbol{\beta})P(\boldsymbol{\varphi})d\boldsymbol{\beta}d\boldsymbol{\varphi}
\end{equation}
This marginal density is analytically integrable over all of the homoscedastic classes defined previously. In heteroscedastic cases (classes VII and VIII), we use a Laplace approximation over the log-variances to approximate $P(\bm{Y}|m_s^{\text{VII}})$ and $P(\bm{Y}|m_s^{\text{VIII}})$. We demonstrate the performance of this approach through a simulation study in Section \ref{sec:ancovasimstudy} and through empirical data sets in Section \ref{ancovaexample}.   

\subsection{Application 2: Unreplicated Two-Way Layouts}\label{sec:twowayms}
Next we consider an unreplicated two-way layout with $R$ rows, $C$ columns, and $N=R\times C$ observations. Because of the unreplicated nature of such a design, the full set of standard interaction effects cannot be incorporated due to insufficient degrees of freedom. Treat the row effects as the SLGF $\boldsymbol{\nu}$, and let $\boldsymbol{\tau}$ contain the column effects; note by transposing the data table we could treat the column effects as the SLGF as well. 


Our choice of model classes is based on the idea of hidden additivity, where interactions are treated as a group-by-column effect \citep{Franck2013}. The usual ``additive" model $\bm{Y}=\bm{1}^T \alpha + W\boldsymbol{\nu}+V\boldsymbol{\tau}+\boldsymbol{\varepsilon}$ accounts for only row and column main effects. The group-based model, partitioned by levels of the row effect, does not include column effects but does consider group-by-column interactions, denoted by $\tilde{\boldsymbol{\rho}}$. We require at least 2 levels of $k$ (rows) in both groups to ensure there are enough degrees of freedom to estimate this group-by-column interaction. We thus begin with the model
\begin{equation}
\bm{Y}=\bm{1}^T \alpha+W\boldsymbol{\nu}+V\boldsymbol{\tau}+U\tilde{\boldsymbol{\rho}}+\boldsymbol{\varepsilon}
\end{equation}
where we let $\tilde{\boldsymbol{\rho}}:=\bm{0}$ in the additive model, $\boldsymbol{\tau}:=\bm{0}$ in cases with scheme-based regression effects, and $\boldsymbol{\varepsilon}:=\tilde{\boldsymbol{\varepsilon}}$ in cases with group-based heteroscedasticity. So the full likelihood function is given by
\begin{equation}
P(\bm{Y}|m_s^c,\alpha, \boldsymbol{\nu}, \boldsymbol{\tau},\tilde{\boldsymbol{\rho}},\boldsymbol{\varphi})=(2\pi)^{-\frac{N}{2}}|\Phi|^{\frac{1}{2}}\exp\{-\frac{1}{2}(Y-\bm{1}\alpha -W\boldsymbol{\nu}-V\boldsymbol{\tau}-U\tilde{\boldsymbol{\rho}})^T \Phi 
(Y-\bm{1}\alpha -W\boldsymbol{\nu}-V\boldsymbol{\tau}-U\tilde{\boldsymbol{\rho}})\}
\end{equation}
In this layout we consider four model classes:
\begin{enumerate}
\item Class I ($m^{\text{I}})$: the "additive" model, where column effects are equivalent across rows and error variance is constant across all observations, contains 1 model with no grouping schemes; 

$\bm{Y}=\bm{1}^T \alpha+W\boldsymbol{\nu}+V\boldsymbol{\tau}+ \boldsymbol{\varepsilon},\  \boldsymbol{\varepsilon}\sim N(\bm{0},\,\sigma^{2}I)$

\item Class II ($m_{s}^{\text{II}})$: the "group-by-column interaction" model, where levels are divided into
two latent groups based on grouping scheme ${d_s}(\bm{k})$ with distinct means and equivalent variance, contains $2^{K-1}-K-1$ grouping schemes \citep{Franck2013}; 

$\bm{Y}=\bm{1}^T \alpha+W\boldsymbol{\nu}+U\tilde{\boldsymbol{\rho}}+\boldsymbol{\varepsilon},\  \boldsymbol{\varepsilon}\sim N(\bm{0},\,\sigma^{2}I)$

\item Class III ($m_{s}^{\text{III}})$: the "heteroscedastic additive" model, where levels are divided into
two latent groups based on grouping scheme ${d_s}(\bm{k})$ with equivalent means and group-based variances, contains $2^{K-1}-K-1$ grouping schemes;

$\bm{Y}=\bm{1}^T \alpha+W\boldsymbol{\nu}+V\boldsymbol{\tau}+ \tilde{\boldsymbol{\varepsilon}},\  \tilde{\boldsymbol{\varepsilon}}\sim N\left(\bm{0},\,\ensuremath{\ensuremath{\tilde{\Sigma}=\left[\begin{array}{c|c}
\sigma_{1}^{2}I_{n_{1}\times n_{1}} & 0_{n_{1}\times n_{2}}\\
\hline 0_{n_{2}\times n_{1}} & \sigma_{2}^{2}I_{n_{2}\times n_{2}}
\end{array}\right]_{N\times N}}}\right)$

\item Class IV ($m_{s}^{\text{IV}})$: the "heteroscedastic group-by-column interaction" model where levels are divided into
two latent groups based on grouping scheme ${d_s}(\bm{k})$ with distinct means and group-based variances, contains $2^{K-1}-K-1$ grouping schemes;

$\bm{Y}=\bm{1}^T \alpha+W\boldsymbol{\nu}+U\tilde{\boldsymbol{\rho}}+ \tilde{\boldsymbol{\varepsilon}},\  \tilde{\boldsymbol{\varepsilon}}\sim N\left(\bm{0},\,\ensuremath{\ensuremath{\tilde{\Sigma}=\left[\begin{array}{c|c}
\sigma_{1}^{2}I_{n_{1}\times n_{1}} & 0_{n_{1}\times n_{2}}\\
\hline 0_{n_{2}\times n_{1}} & \sigma_{2}^{2}I_{n_{2}\times n_{2}}
\end{array}\right]_{N\times N}}}\right)$

\end{enumerate}
Thus there are $M=(3\times2^{K-1})-3K-2$ models considered. 

\subsubsection{Bayesian Model Specification: Unreplicated Two-Way Layouts}\label{sec:twowaybms}
The two-way unreplicated layout is typically modeled with a high ratio of parameters to data points. Thus we must take care in choosing priors that allow us to successfully incorporate the FBF approach. With noninformative priors on the regression effects, the minimal training sample size needed to estimate $R-1$ regression effects, $2\cdot(C-1)$ group-by-column interactions, and $2$ error variances would be prohibitively large in relation to the sample size $N$; indeed, $b=\frac{m_0}{N}<\frac{1}{2}$ only when $R\ge7$. In this work, we propose using the Zellner-Siow mixture $g$-prior on regression coefficients to reduce the dimensionality of the improper prior. Our use of this automatic prior on the regression effects lowers the minimal training sample size to $m_0=3$, and thus the fractional exponent $b=\frac{m_0}{N}$, to a value that allows us to successfully implement the FBF. Thus we let
\begin{equation}
P(\alpha)\propto 1,
\end{equation}
\begin{equation}
P(\boldsymbol{\beta}| \boldsymbol{\varphi},\, g)=N(\bm{0},\\g(X^T \Phi X)^{-1}),
\end{equation}
\begin{equation}
P(\boldsymbol{\varphi})\propto \varphi^{-1}\ \text{(for models with homoscedasticity)},
\end{equation}
\begin{equation}
P(\boldsymbol{\varphi})\propto \varphi_{1}^{-1}\varphi_{2}^{-1}\ \text{(for models with heteroscedasticity)}, \text{ and}
\end{equation}
\begin{equation}
P(g)=\text{IG}\left( \frac{1}{2},\ \frac{N}{2} \right)
\end{equation}
The marginal density of the data conditional on the model is
\begin{equation}
P(\bm{Y}|m_{s}^{c})=\iiiint
P(\bm{Y}|\alpha,\boldsymbol{\beta},\boldsymbol{\varphi},m_{s}^{c})P(\alpha)P(\boldsymbol{\beta}|\boldsymbol{\varphi},g)P(\boldsymbol{\varphi})P(g)d\alpha d\boldsymbol{\beta} d\boldsymbol{\varphi} dg
\end{equation}


It is well-known in the homoscedastic case that the Zellner-Siow mixture $g$-prior advantageously elicits a Cauchy distribution on the regression effects \citep{Liangetal}. We show the Cauchy result also holds in the heteroscedastic case; that is, $\boldsymbol{\beta} \sim \text{MVCauchy}_{p}\left(\text{location}=0,\,\text{scale}=\left(\frac{X^{T}\Phi X}{n}\right)^{-1}\right)$; see online Supplement for proof.

In classes with homoscedasticity, this intergral is intractable over $g$ and thus a Laplace approximation is conducted over a single dimension; in heteroscedastic cases, a three-dimensional Laplace approximation is used to integrate $g$, $\lambda_1=\log(\varphi_1)$, and $\lambda_2=\log(\varphi_2$).

\section{Simulation Studies}
\subsection{Simulation Study: ANCOVA Models}\label{sec:ancovasimstudy}
In order to simulate ANCOVA data, we generated independent draws $x$ uniformly over the interval $(0,10)$. Outcomes $\bm{Y}$ were then simulated according to each of the eight classes described in Section \ref{sec:ancovams} with continuous, categorical, interaction, and/or group-based effects, as well as errors with group-based heteroscedasticity, as appropriate. In this study, we simulated $n_k=90$ observations at each level of the SLGF; another simulation study with $n_k = 10$ is provided in the Supplement. We note that the settings in Table \ref{table:ancovasettings} are not calibrated to be equivalent in the total effect between classes, where such a calibration would be non-trivial. For example, the Class V model contains eight non-null parameters compared to the Class I model's single parameter; thus we avoid comparing overall performance between classes in this study. 
{\small{}}
\begin{table}
\begin{centering}
{\small{}}%
\begin{tabular}{|c|c|}
\hline 
{\small{}Class} & {\small{}Parameters}\tabularnewline
\hline 
\multirow{1}{*}{{\small{}I (Null)}} & {\small{}$\sigma^{2}=1$}\tabularnewline
\hline 
\multirow{1}{*}{{\small{}II (SLR)}} & {\small{}$\tau=0.5,$ $\sigma^{2}=1$}\tabularnewline
\hline 
{\small{}III (ANCOVA)} & {\small{}$\alpha=2.0,\, \nu=(4.0,6.0,8.0)$, $\tau=0.5$, $\sigma^{2}=1$}\tabularnewline
\hline 
{\small{}IV (Group-Contracted ANCOVA)} & {\small{}$\alpha=0.0,\, \tilde{\nu}=(3.0)$, $\tau=0.5$, $\sigma^{2}=1$}\tabularnewline
\hline 
{\small{}V (Interaction ANCOVA)} & {\small{}$\alpha=0.5,\, \nu=(1.0,1.5,2.0)$, $\rho=(0.25,0.5,0.75,1.0)$, $\tau=0.5$,
$\sigma^{2}=1$}\tabularnewline
\hline 
{\small{}VI (Group-Interaction ANCOVA)} & {\small{}$\alpha=0.0,\,\tilde{\nu}=(0.8)$, $\tilde{\rho}=(0,1)$, $\tau=1$,
$\sigma^{2}=1$}\tabularnewline
\hline 
{\small{}VII (Heteroscedastic } & {\small{}$\alpha=0.0,\, \tilde{\nu}=(3.0)$, $\tau=0.5$, }\tabularnewline
{\small{}Group-Contracted ANCOVA)} & {\small{}$\sigma_{1}^{2}=1$, $\sigma_{2}^{2}=5$}\tabularnewline
\hline 
{\small{}VIII (Heteroscedastic } & {\small{}$\alpha=0.0,\, \tilde{\nu}=(3.0)$, $\tau=0.5$, $\tilde{\rho}=(0,1)$, }\tabularnewline
{\small{}Group-Interaction }ANCOVA) & {\small{}$\sigma_{1}^{2}=1$, $\sigma_{2}^{2}=5$}\tabularnewline
\hline 
\end{tabular}
\par\end{centering}{\small \par}
{\small{}\caption{Settings for the eight model classes in the ANCOVA simulation study
where $\alpha=0$, $K=4$, and $N=360$. }\label{table:ancovasettings}
}{\small \par}
\end{table}
{\small \par}
\begin{figure}[htp]
\begin{centering}
\includegraphics[scale=.75]{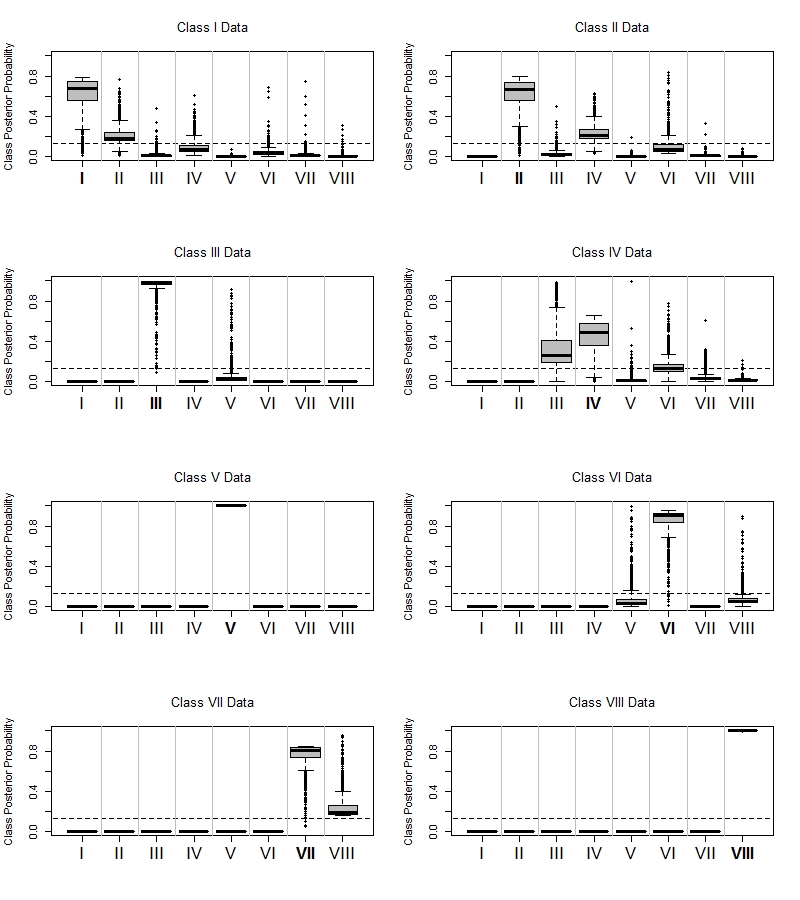}
\caption{Posterior probabilities ($y$-axis) by class based on 1000 Monte Carlo data sets with $K=4$ levels
of the categorical predictor, each with 90 observations, for a total of $N=360$ observations. The true
model class is emphasized in bold on the $x$-axis. The dashed line indicates the prior by model class.}\label{fig:ancovapostprobs}
\end{centering}
\end{figure}

Figure \ref{fig:ancovapostprobs} shows that overall our method accurately attributes posterior model probability to the correct model class for the eight classes described in Section \ref{sec:ancovams}. For Class I (null model) data, the posterior probability of the true class is high relative to the other classes, indicating that our method neither misattributes noise to a group-based structure, nor appears to systematically favor one of the other model classes when it does select the wrong model. Class II (SLR model) data performs similarly, but Classes IV (group-contracted ANCOVA) and VII (group-interaction) capture some posterior probability as well. The Class III (ANCOVA) data setting is favored by the correct class, with Class V (interaction ANCOVA) being the second choice. Class IV (group-contracted ANCOVA) data appears to be the most difficult to detect based on the parameters chosen for this study; although its posterior probabilities are generally higher than the other classes', it is often mistaken for Class III data, meaning in this case a group-based categorical effect is often mistaken for the full categorical effect. Note that these misattributed classes differ by only two parameters in the regression structure. For Class V (interaction ANCOVA) data, the correct model class is favored. Class VI (group-interaction) data is also correctly favored the majority of the time; when misclassified, it is usually chosen as Class V (interaction ANCOVA) or VIII (heteroscedastic ANCOVA). Classes VII (heteroscedastic group-contracted) and VIII (heteroscedastic group-interaction) perform well; when wrong, Class VII tends to favor a spurious interaction effect. 

These results indicate that, in general, our method tends to identify the correct class. When no group-based structure is present, our method tends to not erroneously fit spurious effects or variances. Thus group-based regression effects, interactions, and/or variances are detected with high accuracy when present. We see similar performance with the additional study provided in the online Supplement where $N=40$. 

\subsection{Simulation Study: Unreplicated Two-Way Layouts}
\label{sec:twowaysimstudy}
Row and column effects along with error variance(s) (provided in Table \ref{table:twowaysettings1}) were simulated to generate unreplicated two-way layouts. We consider layouts of size $10\times 5$ where $N=50$. Another setting with a smaller effect size, and a study on layouts of size $5 \times 5$, are given in the Supplement. 
\begin{center}
{\small{}}
\begin{table}
\begin{centering}
{\small{}}%
\begin{tabular}{|>{\centering}p{5cm}|c||c|}
\hline 
{\small{}Class} & \multicolumn{2}{c|}{Parameters}\tabularnewline
\hline 
\multicolumn{1}{|c|}{{\small{}I (Additive Model)}} & \multicolumn{2}{c|}{{\small{}$\alpha=1,\, \boldsymbol{\nu}\in\{2,\,3,\,4,\,5,\,6,\,7,\,8,\,9,\,10\}$,
$\boldsymbol{\tau}\in\{1,\,2,\,3,\,4,\,5\}$, $\sigma^{2}=1$}}\tabularnewline
\hline 
{\small{}II (Group-by-Column } & \multicolumn{2}{c|}{{\small{}$\alpha=1,\, \boldsymbol{\nu}\in\{2,\,3,\,4,\,5,\,6,\,7,\,8,\,9,\,10\}$,
$\boldsymbol{\tau}_{1}\in\{1.0,\,1.8,\,2.6,\,3.4,\,4.2\}$, }}\tabularnewline
{\small{}Interaction)} & \multicolumn{2}{c|}{{\small{}$\boldsymbol{\tau}_{2}\in\{4.2,\,3.4,\,2.6,\,1.8,\,1.0\}$,
$\sigma^{2}=1$}}\tabularnewline
\hline 
{\small{}III (Heteroscedastic } & \multicolumn{2}{c|}{{\small{}$\alpha=1,\, \boldsymbol{\nu}\in\{2,\,3,\,4,\,5,\,6,\,7,\,8,\,9,\,10\}$,
$\boldsymbol{\tau}\in\{1,\,2,\,3,\,4,\,5\}$, }}\tabularnewline
{\small{}Additive)} & \multicolumn{2}{c|}{{\small{}$\sigma_{1}^{2}=1.0$, $\sigma_{2}^{2}=0.10$}}\tabularnewline
\hline 
{\small{}IV (Heteroscedastic Group-} & \multicolumn{2}{c|}{{\small{}$\alpha=1,\, \boldsymbol{\nu}\in\{2,\,3,\,4,\,5,\,6,\,7,\,8,\,9,\,10\}$,
$\boldsymbol{\tau}_{1}\in\{1.0,\,1.8,\,2.6,\,3.4,\,4.2\}$, }}\tabularnewline
{\small{}by-Column Interaction)} & \multicolumn{2}{c|}{{\small{}$\boldsymbol{\tau}_{2}\in\{4.2,\,3.4,\,2.6,\,1.8,\,1.0\}$,
$\sigma_{1}^{2}=1.0$, $\sigma_{2}^{2}=0.10$}}\tabularnewline
\hline 
\end{tabular}
\par\end{centering}{\small \par}
{\small{}\caption{Settings for the four model classes in the two-way unreplicated layout simulation study with $10\times 5$ layouts where $\alpha =0$.}\label{table:twowaysettings1}
}{\small \par}
\end{table}
\par\end{center}{\small \par}
\begin{figure}[htp] 
\begin{centering} 
\includegraphics[scale=.8]{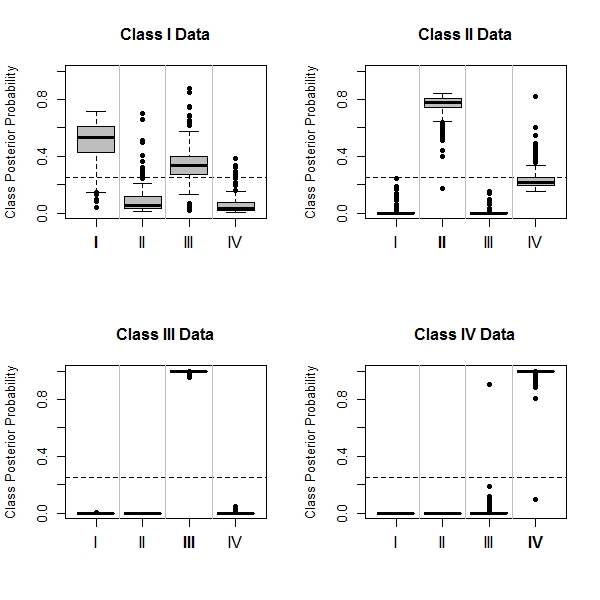}
\par\end{centering}
\caption{Posterior probabilities ($y$-axis) by class based on 1000 Monte Carlo $10\times 5$ layouts. The true model class is emphasized in bold on the $x$-axis. The dashed line indicates the prior by model class.}
\label{fig:twowaypostprobs}
\end{figure}
Figure \ref{fig:twowaypostprobs} shows that our method also tends to favor the true class in the two-way unreplicated layout, with high parameter to data ratio and the mixture $g$-prior. For additive data from Class I, the true class is generally favored; the most probable alternative is typically the heteroscedastic additive Class III. Under large effect size, Class II (group-by-column interaction) data is correctly identified. Class III (heteroscedastic additive) and Class IV (heteroscedastic group-by-column interaction) data are both favored correctly as well regarding both effect sizes. 

\section{Case Studies}
\subsection{Case Study: ANCOVA}\label{ancovaexample}
\begin{figure}[htp] 
\begin{centering}
\includegraphics[width=4.5in]{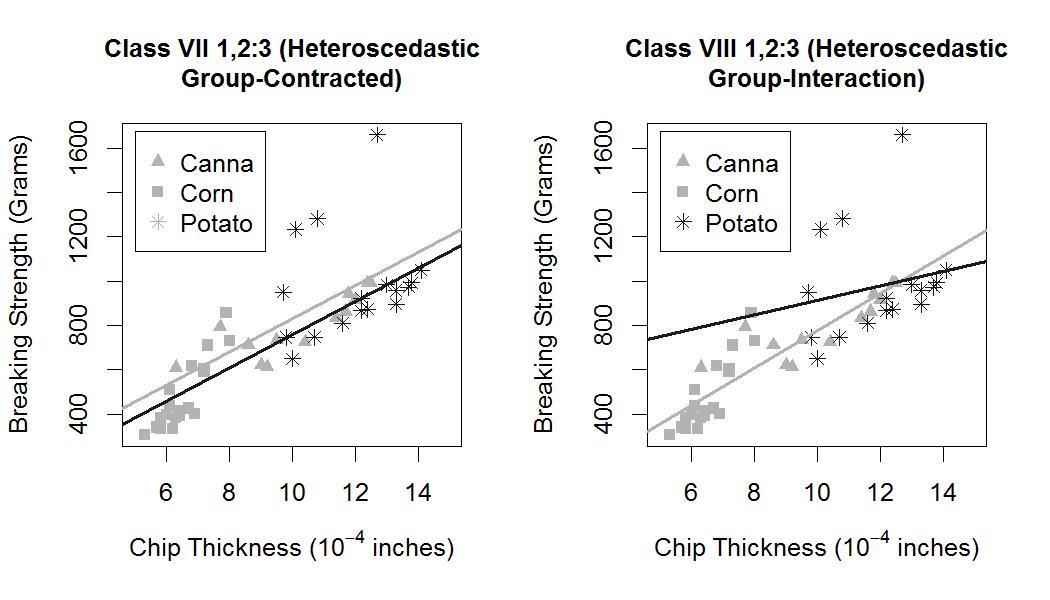}
\par\end{centering}
\caption{The most probable models from the \citet{Flurry} data set are plotted above: the heteroscedastic group-contracted model (left), with $P(m_{1,2:3}^{\text{VII}}|\bm{Y})\approx 0.51$, and the heteroscedastic group-interaction model (right), with $P(m_{1,2:3}^{\text{VIII}}|\bm{Y})\approx 0.42$. Overall the grouping scheme where canna and corn are grouped together accounts for about 93.8\% of the posterior model probability, while heteroscedastic models account for about 93.0\% of the posterior model probability.}
\label{fig:flurry}
\end{figure}

We revisit the \citet{Flurry} data, shown in the center panel of Figure \ref{fig:threeplots} and in Figure \ref{fig:flurry}. The breaking strength of a chip coated with a film derived from one of three plant materials is studied as a function of a continuous predictor (the thickness of the film) and a categorical predictor (the plant type from which the film was developed). The plot seems to indicate some degree of heteroscedasticity between canna and corn, versus potato; results are shown in Figure \ref{fig:flurry}. Two models receive 93\% of the posterior probability: the heteroscedastic group-contracted model, with $P(m_{1,2:3}^{\text{VII}}|\bm{Y})\approx 0.51$, and the heteroscedastic group-interaction model, with $P(m_{1,2:3}^{\text{VIII}}|\bm{Y})\approx 0.42$. 


\subsection{Case Study: Two-Way Unreplicated Layouts}\label{sec:twowaycases}
\begin{figure}[htp]
\begin{centering}
\includegraphics[width=4.5in]{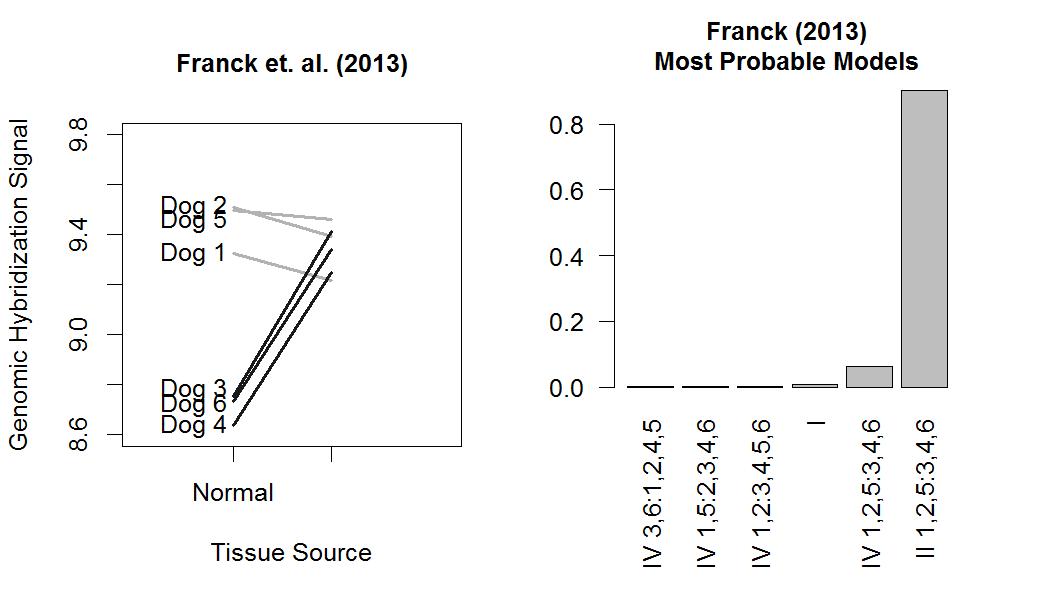}
\par\end{centering}
\caption{A two-way unreplicated layout with apparent group-based regression effects \citep{Franck2013}, where $P(m^{\text{II}}_{1,2,5:3,4,6}|\bm{Y})\approx 0.904$ and $P(m^{\text{IV}}_{1,2,5:3,4,6}|\bm{Y})\approx 0.064$. Overall the grouping scheme $1,2,5:3,4,6$ accounts for approximately 96.8\% of the posterior model probability.}
\label{fig:franckdogs}
\end{figure}

\citet{Franck2013} and \citet{FranckRJ} examine a two-way unreplicated layout describing the copy number variation for genes in six dogs with lymphoma. Samples were taken from healthy and diseased tissue within each dog. It is clear in the plot that dogs behave differently by group: dogs 1, 2, and 5 appear to behave distinctly from dogs 3, 4, and 6. With six subjects by row, there are $1+3\cdot(2^{6-1}-6-1)=76$ candidate models, including the null model and three classes each containing 25 models. Thus we show only the six most probable models, which account for 98.1\% of the posterior probability; these results are shown in Figure \ref{fig:franckdogs}. We conclude with high probability that subjects 1, 2, and 5 behave distinctly from 3, 4, and 6; this scheme over all classes has probability $P(m_{1,2,5:3,4,6}|\bm{Y})\approx .9680$. More specifically, we also conclude homoscedastic behavior along with this grouping scheme, the group-by-column interaction model, with $P(m_{1,2,5:3,4,6}|\bm{Y})\approx .9040$. 


\section{Discussion}
Our proposed method is a flexible and intuitive approach to accommodate linear models with latent group-based effects underlying the data. This method generalizes the homoscedastic, two-way unreplicated layout work of \citet{Franck2018} to a heteroscedastic approach to any linear model with a categorical predictor. By partitioning the data according to the levels of a categorical predictor, we can detect latent structures within the data that might influence the regression effects, error variance, or interaction effects. Often, these group-based structures have straightforward and intuitive interpretations in the context of the data, making the approach particularly useful to domain experts. The use of fractional Bayes factors allows us to compare homoscedastic and heteroscedastic models with minimal prior influence. Regarding priors on the regression effects, we explored the performance of noninformative and mixture $g$-priors in the contexts of ANCOVA and two-way unreplicated layouts, but many other choices on priors and data layouts can be considered.  We considered cases in which the number of levels of the SLGF $K$ was relatively small, so that a combinatoric search over all possible grouping schemes was computationally feasible. In cases where $K$ is large, a Markov-chain model composition (MC3) could be used to search the model space. 

In some cases, our method leads to a contraction of the number of effects modeled. For instance, in the ANCOVA examples illustrated in Section \ref{sec:ancovams}, estimating a group effect $\tilde{\boldsymbol{\nu}}$ as opposed to a categorical effect $\boldsymbol{\nu}$ will reduce the degrees of freedom used to estimate the effect from $K$ to $2$. Alternatively, in the two-way unreplicated layout examples, modeling a group-by-column interaction rather than column effects will expand the degrees of freedom used from $C-1$ to $2\cdot (C-1)$. 

While we have illustrated our method using two latent groups, an extension to three or more groups is straightforward conceptually but increases the complexity of the models, the number of model classes, and the number of models to consider. Fortunately, the two-group assumption has been shown to lead to useful inferences in several previous works, including \citet{KKSA}, \citet{Franck2013}, \citet{FranckRJ}, and \citet{Franck2018}. 


\section{Online Supplement}

\subsection{Marginal Model Probability Calculations}

To prevent numeric underflow we consider model probabilites on the logarithmic scale in all cases. Consider the set of all log-marginal probabilities $\mathcal{L}=\{\log P(Y|m_{s}^{c})\}$. Let $\ell^{\star}=\max\mathcal{L}$ and $m^{\star}=\underset{m}{\arg\max}\ \mathcal{L}$. Transform $\mathcal{L}^{\star}=\mathcal{L}-\ell^{\star}$ to obtain the set $\exp[\mathcal{L}^{\star}]=\{\exp[\log P(Y|m_{s}^{c})-\log P(Y|m^{\star})]\}=\lbrace\frac{P(Y|m_{s}^{c})}{P(Y|m^{\star})}\rbrace=\lbrace B^{m_{s}^{c},\,m^{\star}}\rbrace$, representing Bayes factors for each model relative to the most probable model. Note the untransformed model probabilities are given by $P=\lbrace\frac{P(Y|m_{s}^{c})P(m_{s}^{c})}{\sum P(Y|m_{s}^{c})P(m_{s}^{c})}\rbrace=\lbrace\frac{\nicefrac{P(Y|m_{s}^{c})P(m_{s}^{c})}{P(Y|m^{\star})}}{\nicefrac{\sum P(Y|m_{s}^{c})P(m_{s}^{c})}{P(Y|m^{\star})}}\rbrace=\lbrace\frac{B^{m_{s}^{c},\,m^{\star}}P(m_{s}^{c})}{\sum B^{m_{s}^{c},\,m^{\star}}P(m_{s}^{c})}\rbrace$;
thus we can easily use these transformed log-marginal probabilities to obtain posterior model probabilities. Our Laplace approximations generally provided values comparable to quadrature and cubature-based approximations at less computational expense. 

\subsection{Derivation of Model Probabilities: Noninformative Regression Effect Priors}

Let $H_X:=X(X^TX)^{-1}X^T$.

\subsubsection{Homoscedastic Error Variance, No Group-Based Regression Effects}

\begin{align*}
P(\bm{Y}|m) & =\underset{\Theta}{\int}\left[(2\pi)^{-\frac{N}{2}}|\Sigma|^{-\frac{1}{2}}\cdot\exp\{-\frac{1}{2}(\bm{Y}-X\bm{\beta})^{T}\Sigma{}^{-1}(\bm{Y}-X\bm{\beta})\}\right]^{b}P(\boldsymbol{\theta})d\boldsymbol{\theta}\\
 & =\underset{0}{\overset{\infty}{\int}}\underset{-\infty}{\overset{\infty}{\int}}(2\pi)^{^{-\frac{Nb}{2}}}|\Sigma|^{-\frac{b}{2}}\exp\{-b\cdot\frac{1}{2}[\bm{Y}^{T}\Sigma{}^{-1}\bm{Y}-2\bm{\beta}^{T}X^{T}\Sigma^{-1}\bm{Y}+\bm{\beta}^{T}X^{T}\Sigma^{-1}X\bm{\beta}]\}\Sigma^{-1}d\bm{\beta}d\gamma\\
 & =\underset{0}{\overset{\infty}{\int}}\underset{-\infty}{\overset{\infty}{\int}}(2\pi)^{^{-\frac{Nb}{2}}}\gamma{}^{\frac{Nb}{2}-1}\exp\{-b\cdot\frac{1}{2}[\bm{Y}^{T}\Sigma{}^{-1}\bm{Y}-2\bm{\beta}^{T}X^{T}\Sigma^{-1}\bm{Y}+\bm{\beta}^{T}X^{T}\Sigma^{-1}X\bm{\beta}]\}d\bm{\beta}d\gamma\\
 & =\underset{0}{\overset{\infty}{\int}}\underset{-\infty}{\overset{\infty}{\int}}(2\pi)^{^{-\frac{Nb}{2}}}\gamma{}^{\frac{Nb}{2}-1}\exp\{-b\cdot\frac{\gamma}{2}[\bm{\beta}^{T}X^{T}X\bm{\beta}-2\bm{\beta}^{T}X^{T}\bm{Y}+\bm{Y}^{T}H\bm{Y}]\}\times\\
 & \qquad\qquad\exp\{-b\cdot\frac{\gamma}{2}[\bm{Y}^{T}\bm{Y}-\bm{Y}^{T}H\bm{Y}]\}d\bm{\beta}d\gamma\\
 & =\underset{0}{\overset{\infty}{\int}}(2\pi)^{^{-\frac{Nb}{2}}}\gamma{}^{\frac{Nb}{2}-1}(2\pi)^{+\frac{p}{2}}|b^{-1}\gamma^{-1}(X^{T}X)^{-1}|^{+\frac{1}{2}}\exp\{-b\cdot\frac{\gamma}{2}[\bm{Y}^{T}(I-H)\bm{Y}]\}d\gamma\\
 & =(2\pi)^{-\frac{Nb-P}{2}}b^{-\frac{P}{2}}|X^{T}X|^{-\frac{1}{2}}\Gamma\left(\frac{Nb-P}{2}\right)\left(\frac{b\cdot\text{SSResid}^{\text{I}}}{2}\right)^{-\frac{Nb-P}{2}}\\
 & =\pi^{-\frac{Nb-P}{2}}b^{-\frac{Nb}{2}}|X^{T}X|^{-\frac{1}{2}}\Gamma\left(\frac{Nb-P}{2}\right)(\text{SSResid}^{\text{I}})^{-\frac{Nb-P}{2}}.\ \square
\end{align*}

\begin{align*}
q^{b}(\boldsymbol{Y}|m) & =\frac{\underset{\Theta}{\int}P(Y|m^{\text{I}},\boldsymbol{\theta})P(\boldsymbol{\theta})d\boldsymbol{\theta}}{\underset{\Theta}{\int}P^b(Y|m^{\text{I}},\boldsymbol{\theta})P(\boldsymbol{\theta})d\boldsymbol{\theta}}\\
 & =\frac{\pi^{-\frac{N-P}{2}}|X^{T}X|^{-\frac{1}{2}}\Gamma\left(\frac{N-P}{2}\right)(\text{SSResid}^{\text{I}})^{-\frac{N-P}{2}}.}{\pi^{-\frac{Nb-P}{2}}b^{-\frac{Nb}{2}}|X^{T}X|^{-\frac{1}{2}}\Gamma\left(\frac{Nb-P}{2}\right)(\text{SSResid}^{\text{I}})^{-\frac{Nb-P}{2}}}\\
 & =\pi^{-\frac{N(1-b)}{2}}b^{\frac{Nb}{2}}(\text{SSResid}^{\text{I}})^{-\frac{N(1-b)}{2}}\frac{\Gamma\left(\frac{N-P}{2}\right)}{\Gamma\left(\frac{Nb-P}{2}\right)}.\ \square
\end{align*}

\subsubsection{Homoscedastic Error Variance, Group-Based Regression Effects}

\begin{align*}
P(\bm{Y}|m_{s}) & =\underset{\Theta}{\int}\left[(2\pi)^{-\frac{N}{2}}|\Sigma|^{-\frac{1}{2}}\cdot\exp\{-\frac{1}{2}(\bm{Y}-X\bm{\beta})^{T}\Sigma{}^{-1}(\bm{Y}-X\bm{\beta})\}\right]^{b}P(\boldsymbol{\theta})d\boldsymbol{\theta}\\
 & =\underset{0}{\overset{\infty}{\int}}\underset{-\infty}{\overset{\infty}{\int}}\underset{-\infty}{\overset{\infty}{\int}}(2\pi)^{^{-\frac{Nb}{2}}}\gamma{}^{\frac{Nb}{2}-1}\exp\{-b\cdot\frac{1}{2}[\bm{Y}^{T}\Sigma{}^{-1}\bm{Y}-2\boldsymbol{\beta}^{T}X^{T}\Sigma^{-1}\bm{Y}+\boldsymbol{\beta}^{T}X^{T}\Sigma^{-1}X\boldsymbol{\beta}^{T}]\}d\boldsymbol{\beta}_{1}^{T}d\boldsymbol{\beta}_{2}^{T}d\gamma\\
 & =\underset{0}{\overset{\infty}{\int}}\underset{-\infty}{\overset{\infty}{\int}}\underset{-\infty}{\overset{\infty}{\int}}(2\pi)^{^{-\frac{Nb}{2}}}\gamma{}^{\frac{Nb}{2}-1}\exp\{-b\cdot\frac{\gamma}{2}[\boldsymbol{\beta}^{T}X^{T}X\boldsymbol{\beta}-2\boldsymbol{\beta}^{T}X^{T}\boldsymbol{Y}+\boldsymbol{Y}^{T}H\boldsymbol{Y}]\}\times\\
 & \qquad\qquad\exp\{-b\cdot\frac{\gamma}{2}[\boldsymbol{Y}^{T}\boldsymbol{Y}-\boldsymbol{Y}^{T}H\boldsymbol{Y}]\}d\boldsymbol{\beta}_{1}d\boldsymbol{\beta}_{2}d\gamma\\
 & =\underset{0}{\overset{\infty}{\int}}(2\pi)^{^{-\frac{Nb}{2}}}\gamma{}^{\frac{Nb}{2}-1}(2\pi)^{+\frac{P}{2}}|b^{-1}\gamma^{-1}(X_{1}X_{1})^{-1}|^{+\frac{1}{2}}|b^{-1}\gamma^{-1}(X_{2}X_{2})^{-1}|^{+\frac{1}{2}}\times\\
 & \qquad\qquad\exp\{-b\cdot\frac{\gamma}{2}[\boldsymbol{Y}_{1}^{T}(I-H_{1})\boldsymbol{Y}_{1}+\boldsymbol{Y}_{2}^{T}(I-H_{2})\boldsymbol{Y}_{2}]\}d\gamma\\
 & =(2\pi)^{-\frac{Nb-P}{2}}b^{-\frac{P}{2}}|X_{1}^{T}X_{1}|^{-\frac{1}{2}}|X_{2}^{T}X_{2}|^{-\frac{1}{2}}\Gamma\left(\frac{Nb-P}{2}\right)\left(\frac{b\cdot\left[\text{SSResid}_{1}^{\text{II}}+\text{SSResid}_{2}^{\text{II}}\right]}{2}\right)^{-\frac{Nb-P}{2}}\\
 & =\pi^{-\frac{Nb-P}{2}}b^{-\frac{Nb}{2}}|X_{1}^{T}X_{1}|^{-\frac{1}{2}}|X_{2}^{T}X_{2}|^{-\frac{1}{2}}\Gamma\left(\frac{Nb-P}{2}\right)(\text{SSResid}_{1}^{\text{II}}+\text{SSResid}_{2}^{\text{II}})^{-\frac{Nb-P}{2}}.\ \square
\end{align*}

\begin{align*}
q^{b}(\boldsymbol{Y}|m_{s}) & =\frac{\underset{\Theta}{\int}P(Y|m_{s}^{\text{II}},\boldsymbol{\theta})P(\boldsymbol{\theta})d\boldsymbol{\theta}}{\underset{\Theta}{\int}P^b(Y|m_{s}^{\text{II}},\boldsymbol{\theta})P(\boldsymbol{\theta})d\boldsymbol{\theta}}\\
 & =\frac{\pi^{-\frac{N-P}{2}}b^{-\frac{N}{2}}|X_{1}^{T}X_{1}|^{-\frac{1}{2}}|X_{2}^{T}X_{2}|^{-\frac{1}{2}}\Gamma\left(\frac{N-P}{2}\right)(\text{SSResid}_{1}^{\text{II}}+\text{SSResid}_{2}^{\text{II}})^{-\frac{N-P}{2}}}{\pi^{-\frac{Nb-P}{2}}b^{-\frac{Nb}{2}}|X_{1}^{T}X_{1}|^{-\frac{1}{2}}|X_{2}^{T}X_{2}|^{-\frac{1}{2}}\Gamma\left(\frac{Nb-P}{2}\right)(\text{SSResid}_{1}^{\text{II}}+\text{SSResid}_{2}^{\text{II}})^{-\frac{Nb-P}{2}}}\\
 & =\pi^{-\frac{N(1-b)}{2}}b^{\frac{Nb}{2}}(\text{SSResid}_{1}^{\text{II}}+\text{SSResid}_{2}^{\text{II}})^{-\frac{N(1-b)}{2}}\frac{\Gamma\left(\frac{N-P}{2}\right)}{\Gamma\left(\frac{Nb-P}{2}\right)}.\ \square
\end{align*}

\subsubsection{Heteroscedastic Error Variance}

A Laplace approximation is used to evaluate $\int P(\bm{Y}|\boldsymbol{\varphi},m)P(\boldsymbol{\varphi})d\boldsymbol{\varphi}$, parametrized with respect to the log-variances $\lambda_1=\log{\sigma^2_1}$ and $\lambda_2=\log{\sigma^2_2}$. Denote $\Lambda$ as the log-variance matrix, $J_{\Lambda}$ as the transformation Jacobian, $(\lambda_1^\star,\,\lambda_2^\star)$ as the mode of the log-marginal distribution, and $\nabla^{\star}$ as the Hessian evaluated at this mode. A subscript of $b$ refers to the same quantities calculated with respect to the fractional exponentiated likelihood. The joint modes $(\lambda_1^\star,\,\lambda_2^\star)$ and $(\lambda_{b_1}^\star,\,\lambda_{b_2}^\star)$ are computed using the function \texttt{optim} in R; similarly, the Hessians $\nabla^\star$ and $\nabla_b^\star$ are evaluated at these values using the function \texttt{hessian} in the package \texttt{numderiv} \citep{numderiv, statspackage}. 

We first integrate over the regression effects vector $\boldsymbol{\beta}$. Let $H_{\Phi}:=\Phi X(X^T\Phi X)^{-1}X^T\Phi$.

\begin{align*}
P^{b}(\bm{Y}|m_{s}^{\text{III}},\varphi_{1},\varphi_{2}) & =\underset{\Theta}{\int}\left[(2\pi)^{-\frac{N}{2}}|\Phi|^{\frac{1}{2}}\cdot\exp\{-\frac{1}{2}(\bm{Y}-X\bm{\beta})^{T}\Phi(\bm{Y}-X\bm{\beta})\}\right]^{b}P(\boldsymbol{\theta})d\boldsymbol{\theta}\\
 & =\underset{-\infty}{\overset{\infty}{\int}}(2\pi)^{^{-\frac{Nb}{2}}}\varphi_{1}{}^{\frac{n_{1}b}{2}-1}\varphi_{2}{}^{\frac{n_{2}b}{2}-1}\exp\{-b\cdot\frac{1}{2}[\bm{Y}^{T}\Phi\bm{Y}-2\boldsymbol{\beta}^{T}X^{T}\Phi\bm{Y}+\boldsymbol{\beta}^{T}X^{T}\Phi X\boldsymbol{\beta}^{T}]\}d\boldsymbol{\beta}\\
 & =\underset{-\infty}{\overset{\infty}{\int}}(2\pi)^{^{-\frac{Nb}{2}}}\varphi_{1}{}^{\frac{n_{1}b}{2}-1}\varphi_{2}{}^{\frac{n_{2}b}{2}-1}\times\\
 & \qquad\qquad\exp\{-b\cdot\frac{1}{2}[\boldsymbol{\beta}^{T}X^{T}\Phi X\boldsymbol{\beta}-2\boldsymbol{\beta}^{T}X^{T}\boldsymbol{Y}+\boldsymbol{Y}^{T}H_{\Phi}\boldsymbol{Y}]\}\times\\
 & \qquad\qquad\exp\{-b\cdot\frac{1}{2}[\boldsymbol{Y}^{T}\Phi\boldsymbol{Y}-\boldsymbol{Y}^{T}H_{\Phi}\boldsymbol{Y}]\}d\boldsymbol{\beta}\\
 & =(2\pi)^{-\frac{Nb-P}{2}}\varphi_{1}{}^{\frac{n_{1}b}{2}-1}\varphi_{2}{}^{\frac{n_{2}b}{2}-1}b^{-\frac{P}{2}}|X^{T}\Phi X|^{-\frac{1}{2}}\exp\{-b\cdot\frac{1}{2}[\bm{Y}^{T}\Phi\bm{Y}-\bm{Y}^{T}H_{\Phi}\bm{Y}]\}
\end{align*}

We reparametrize the precisions of $P^b(\bm{Y}|m_s^{\text{III}},\varphi_1,\varphi_2)$ and $P(\bm{Y}|m_s^{\text{III}},\varphi_1,\varphi_2)$ to log-variances $\lambda_1$ and $\lambda_2$ to elicit a shape more conducive to the Laplace approximation. 

\begin{figure}[htp] 
\begin{centering}
\includegraphics[scale=0.50]{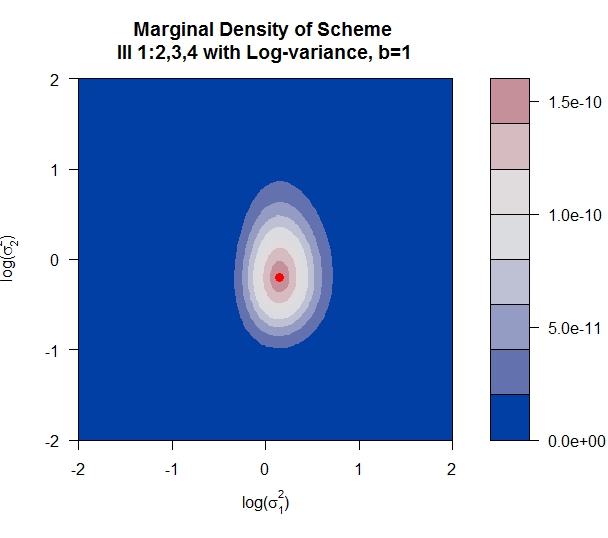} \includegraphics[scale=0.50]{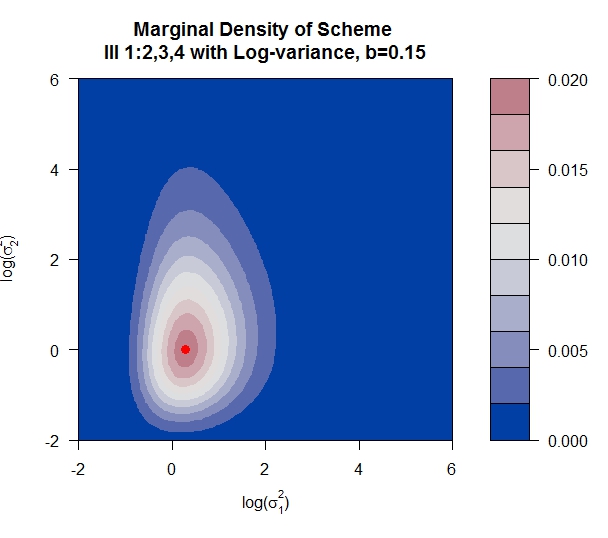}
\par\end{centering}
\caption{The marginal densities $P(\bm{Y}|m^{\text{II}}_{1:2,3,4},\lambda_1,\lambda_2)$ and $P^b(\bm{Y}|m^{\text{III}}_{1:2,3,4},\lambda_1,\lambda_2)$ for one particular grouping of an ANOVA layout with $K=4$ levels of $n_k=10$ observations each show an overall shape conducive to the Laplace approximation in both the raw and exponentiated likelihood cases.}
\end{figure}

For $\lambda_1=\ln{\varphi_1^{-1}}$ and $\lambda_2=\ln{\varphi_2^{-1}}$, $P(\bm{Y}|m_s^{\text{III}},\lambda_1,\lambda_2)=P(\bm{Y}|m_s^{\text{III}},\varphi_1,\varphi_2)\cdot|J_{\Lambda}|$ where $\frac{\partial \varphi_1}{\partial \lambda_1}=-\exp\{-\lambda_1\}$ and $\frac{\partial \varphi_2}{\partial \lambda_2}=-\exp\{-\lambda_2\}$, so $|J_{\Lambda}|=\exp\{-(\lambda_1+\lambda_2)\}$. 

Thus the Laplace approximation for the fractional marginal model probability is given by 

\begin{align*}
q^b(\bm{Y}|m_s^{\text{III}})\approx \frac{(2\pi)|-\nabla^{\star}|^{-\frac{1}{2}}\cdot P(\bm{Y}|m_s^{\text{III}},\varphi_{b_{1}}^{\star},\varphi_{b_{2}}^{\star})|J_{\Lambda}^{\star}|}{(2\pi)|-\nabla_b^{\star}|^{-\frac{1}{2}}\cdot P^b(\bm{Y}|m_s^{\text{III}},\varphi_{b_{1}}^{\star},\varphi_{b_{2}}^{\star})|J_{b_{\Lambda}}^{\star}|}.\ \square
\end{align*}

Empirical comparisons with cubature-based approximations, in addition to the shapes of the marginal densities $P(\bm{Y}|m_s^{\text{III}},\lambda_1,\lambda_2)$ and $P^b(\bm{Y}|m_s^{\text{III}},\lambda_1,\lambda_2)$, indicate that this Laplace approximation is satisfactory.  

\subsubsection{Special Case: Heteroscedastic Error Variance with no Regression Effects Spanning Two Variances}

\begin{align*}
P(\bm{Y}|m_{s}) & =\underset{\Theta}{\int}\left[(2\pi)^{-\frac{N}{2}}|\Sigma|^{-\frac{1}{2}}\cdot\exp\{-\frac{1}{2}(\boldsymbol{Y}-X\boldsymbol{\beta})^{T}\Sigma{}^{-1}(\boldsymbol{Y}-X\boldsymbol{\beta})\}\right]^{b}P(\boldsymbol{\theta})d\boldsymbol{\theta}\\
 & =\underset{0}{\overset{\infty}{\int}}\underset{0}{\overset{\infty}{\int}}\underset{-\infty}{\overset{\infty}{\int}}\underset{-\infty}{\overset{\infty}{\int}}(2\pi)^{^{-\frac{Nb}{2}}}|\Sigma|^{-\frac{b}{2}}\exp\{-b\cdot\frac{1}{2}[\boldsymbol{Y}^{T}\Sigma{}^{-1}\boldsymbol{Y}-2\boldsymbol{\beta}^{T}X^{T}\Sigma^{-1}\boldsymbol{Y}+\boldsymbol{\beta}^{T}X^{T}\Sigma^{-1}X\boldsymbol{\beta}]\}\\
 & \qquad\qquad d\boldsymbol{\beta}_{1}d\boldsymbol{\beta}_{2}d\gamma_{1}d\gamma_{2}\\
 & =\underset{0}{\overset{\infty}{\int}}\underset{0}{\overset{\infty}{\int}}\underset{-\infty}{\overset{\infty}{\int}}\underset{-\infty}{\overset{\infty}{\int}}(2\pi)^{^{-\frac{Nb}{2}}}\gamma{}^{\frac{n_{1}b}{2}-1}\gamma{}^{\frac{n_{2}b}{2}-1}\exp\{-b\cdot\frac{1}{2}[\boldsymbol{Y}^{T}\Sigma{}^{-1}\boldsymbol{Y}-2\boldsymbol{\beta}^{T}X^{T}\Sigma^{-1}\boldsymbol{Y}+\boldsymbol{\beta}^{T}X^{T}\Sigma^{-1}X\boldsymbol{\beta}]\}\\
 & \qquad\qquad d\boldsymbol{\beta}_{1}d\boldsymbol{\beta}_{2}d\gamma_{1}d\gamma_{2}\\
 & =\underset{0}{\overset{\infty}{\int}}\underset{0}{\overset{\infty}{\int}}(2\pi)^{^{-\frac{Nb}{2}}}\gamma{}^{\frac{n_{1}b}{2}-1}\gamma{}^{\frac{n_{2}b}{2}-1}(2\pi)^{+\frac{P}{2}}|b^{-1}\gamma^{-1}(X_{1}^{T}X_{1})^{-1}|^{+\frac{1}{2}}|b^{-1}\gamma^{-1}(X_{2}^{T}X_{2})^{-1}|^{+\frac{1}{2}}\times\\
 & \qquad\qquad\exp\{-b\cdot\frac{\gamma}{2}[\boldsymbol{Y}_{1}^{T}(I-H_{1})\boldsymbol{Y}_{1}+\boldsymbol{Y}_{2}^{T}(I-H_{2})\boldsymbol{Y}_{2}]\}d\gamma_{1}d\gamma_{2}\\
 & =(2\pi)^{-\frac{Nb-P}{2}}b^{-\frac{P}{2}}|X_{1}^{T}X_{1}|^{-\frac{1}{2}}|X_{2}^{T}X_{2}|^{-\frac{1}{2}}\Gamma\left(\frac{Nb-P}{2}\right)\left(\frac{b\cdot\left[\text{SSResid}_{1}^{\text{IV}}+\text{SSResid}_{2}^{\text{IV}}\right]}{2}\right)^{-\frac{Nb-P}{2}}\\
 & =(2\pi)^{-\frac{Nb-P}{2}}b^{-\frac{P}{2}}|X_{1}^{T}X_{1}|^{-\frac{1}{2}}|X_{2}^{T}X_{2}|^{-\frac{1}{2}}\Gamma\left(\frac{n_{1}b-p_{1}}{2}\right)\Gamma\left(\frac{n_{2}b-p_{2}}{2}\right)\times\\
 & \qquad\qquad\left(\frac{b\cdot\text{SSResid}_{1}^{\text{IV}}}{2}\right)^{-\frac{n_{1}b-p_{1}}{2}}\left(\frac{b\cdot\text{SSResid}_{2}^{\text{IV}}}{2}\right)^{-\frac{n_{2}b-p_{2}}{2}}.
\end{align*}

\begin{align*}
q^{b}(\boldsymbol{Y}|m_{s}) & =\frac{\underset{\Theta}{\int}P(Y|m_{s}^{},\boldsymbol{\theta})P(\boldsymbol{\theta})d\boldsymbol{\theta}}{\underset{\Theta}{\int}P^b(Y|m_{s}^{\text{IV}},\boldsymbol{\theta})P(\boldsymbol{\theta})d\boldsymbol{\theta}}\\
 & =\frac{(2\pi)^{-\frac{N-P}{2}}|X_{1}X_{1}|^{-\frac{1}{2}}|X_{2}X_{2}|^{-\frac{1}{2}}\Gamma\left(\frac{n_{1}-p_{1}}{2}\right)\Gamma\left(\frac{n_{2}-p_{2}}{2}\right)}{(2\pi)^{-\frac{Nb-P}{2}}b^{-\frac{P}{2}}|X_{1}X_{1}|^{-\frac{1}{2}}|X_{2}X_{2}|^{-\frac{1}{2}}\Gamma\left(\frac{n_{1}b-p_{1}}{2}\right)\Gamma\left(\frac{n_{2}b-p_{2}}{2}\right)}\times\\
 & \qquad\qquad\frac{\left(\frac{\text{SSResid}_{1}^{\text{IV}}}{2}\right)^{-\frac{n_{1}-p_{1}}{2}}\left(\frac{\text{SSResid}_{2}^{\text{IV}}}{2}\right)^{-\frac{n_{2}-p_{2}}{2}}}{\left(\frac{b\cdot\text{SSResid}_{1}^{\text{IV}}}{2}\right)^{-\frac{n_{1}b-p_{1}}{2}}\left(\frac{b\cdot\text{SSResid}_{2}^{\text{IV}}}{2}\right)^{-\frac{n_{2}b-p_{2}}{2}}}\\
 & =\pi^{-\frac{N(1-b)}{2}}b^{\frac{Nb}{2}}(\text{SSResid}_{1}^{\text{IV}})^{-\frac{n_{1}(1-b)}{2}}(\text{SSResid}_{2}^{\text{IV}})^{-\frac{n_{2}(1-b)}{2}}\frac{\Gamma\left(\frac{n_{1}-p_{1}}{2}\right)}{\Gamma\left(\frac{n_{1}b-p_{1}}{2}\right)}\frac{\Gamma\left(\frac{n_{2}-p_{2}}{2}\right)}{\Gamma\left(\frac{n_{2}b-p_{2}}{2}\right)}.\ \square
\end{align*}

\subsection{Derivation of Model Probabilities: Mixture $\bm{g}$ Regression Effect Priors}

\begin{align*}
P(\bm{Y},g|m_{s}) & =\overset{\infty}{\underset{0}{\int}}\overset{\infty}{\underset{0}{\int}}\overset{\infty}{\underset{-\infty}{\int}}\overset{\infty}{\underset{-\infty}{\int}}P^{b}(\bm{Y}|\alpha,\beta,\varphi,m_{s})P(\alpha,\varphi)P(\beta|\varphi,g)P(g)d\alpha d\beta d\varphi dg\\
 & =\overset{\infty}{\underset{0}{\int}}\overset{\infty}{\underset{0}{\int}}\overset{\infty}{\underset{-\infty}{\int}}\overset{\infty}{\underset{-\infty}{\int}}(2\pi)^{-\frac{Nb}{2}}\varphi^{\frac{Nb}{2}}\exp\{-\frac{\varphi b}{2}(\bm{Y}-\bm{1}^{T}\alpha-X\boldsymbol{\beta})^{T}(\bm{Y}-\bm{1}^{T}\alpha-X\boldsymbol{\beta})\}\times\\
 & (2\pi)^{-\frac{P}{2}}|g\varphi^{-1}(X^{T}X)^{-1}|^{-\frac{1}{2}}\exp\{-\frac{\varphi}{2}\frac{\boldsymbol{\beta}^{T}X^{T}X\boldsymbol{\beta}}{g}\}\varphi^{-1}P(g)d\alpha d\boldsymbol{\beta}d\varphi dg\\
 & =\overset{\infty}{\underset{0}{\int}}\overset{\infty}{\underset{0}{\int}}\overset{\infty}{\underset{-\infty}{\int}}\overset{\infty}{\underset{-\infty}{\int}}(2\pi)^{-\frac{Nb+P}{2}}\varphi^{\frac{Nb}{2}-1}\times \\
 & \exp\{-\frac{\varphi}{2}(\alpha^{2}b\bm{1}^{T}\bm{1}-2\alpha b(\bm{1}^{T}\bm{Y}-\bm{1}^{T}X\boldsymbol{\beta})+b(\bm{Y}-X\boldsymbol{\beta})H_{\bm{1}}(\bm{Y}-X\boldsymbol{\beta}))\}\times\\
 & |g\varphi^{-1}(X^{T}X)^{-1}|^{-\frac{1}{2}}\exp\{-\frac{\varphi}{2}(\boldsymbol{\beta}^{T}\left(\frac{X^{T}X}{g}+bX^{T}X\right)\boldsymbol{\beta}-2\boldsymbol{\beta}^{T}bX^{T}\bm{Y})\}\times\\
 & \exp\{-\frac{\varphi}{2}(b\bm{Y}^{T}\bm{Y}-b(\bm{Y}-X\boldsymbol{\beta})H_{\bm{1}}(\bm{Y}-X\boldsymbol{\beta}))\}P(g)d\alpha d\boldsymbol{\beta}d\varphi dg\\
 & =\overset{\infty}{\underset{0}{\int}}\overset{\infty}{\underset{0}{\int}}\overset{\infty}{\underset{-\infty}{\int}}(2\pi)^{-\frac{Nb+P-1}{2}}\varphi^{\frac{Nb+P-1}{2}-1}b^{-\frac{1}{2}}|X^{T}X|^{-\frac{1}{2}}N^{-\frac{1}{2}}g^{-\frac{P}{2}}\times\\
 & \exp\{-\frac{\varphi}{2}(\boldsymbol{\beta}^{T}\left(\frac{X^{T}X}{g}+bX^{T}X\right)\boldsymbol{\beta}-2\boldsymbol{\beta}^{T}bX^{T}\bm{Y}+\frac{b^{2}g}{1+bg}\bm{Y}^{T}H_{X}\bm{Y})\}\times\\
 & \exp\{-\frac{\varphi}{2}(b\bm{Y}^{T}\bm{Y}-b\bm{Y}^{T}H_{\bm{1}}\bm{Y}-\frac{b^{2}g}{1+bg}\bm{Y}^{T}H_{X}\bm{Y})\}P(g)d\boldsymbol{\beta}d\varphi dg\\
 & =\overset{\infty}{\underset{0}{\int}}\overset{\infty}{\underset{0}{\int}}(2\pi)^{-\frac{Nb-1}{2}}\varphi^{\frac{Nb-1}{2}-1}b^{-\frac{1}{2}}N^{-\frac{1}{2}}(1+bg)^{-\frac{P}{2}}\times\\
 & \exp\{-\frac{\varphi}{2}(b\bm{Y}^{T}\bm{Y}-b\bm{Y}^{T}H_{\bm{1}}\bm{Y}-\frac{b^{2}g}{1+bg}\bm{Y}^{T}H_{X}\bm{Y})\}d\varphi dg\\
 & =\overset{\infty}{\underset{0}{\int}}\frac{\Gamma\left(\frac{Nb-1}{2}\right)}{\sqrt{\pi}^{Nb-1}\sqrt{N}}b^{-\frac{Nb}{2}}(1+bg)^{-\frac{Nb-P-1}{2}}\text{SST}^{-\frac{Nb-1}{2}}[1+bg(1-R^{2})]^{-\frac{Nb-1}{2}}P(g)dg.\ \square
\end{align*}

After integrating over $\alpha,\,\boldsymbol{\beta}$, and $\boldsymbol{\varphi}$, we obtain

\begin{equation}
P(\bm{Y},g|m_s^c)=\overset{\infty}{\underset{0}{\int}}\frac{\Gamma(\frac{Nb-1}{2})b^{-\frac{Nb}{2}}}{\sqrt{\pi}^{Nb-1}\sqrt{N}}\frac{\text{SST}^{-\frac{Nb-1}{2}}}{(1+bg)^{\frac{Nb-P-1}{2}}}[1+bg(1-R^{2})]^{-\frac{Nb-1}{2}}g^{-1.5}\exp\{-\frac{N}{2g}\}dg
\end{equation}

To execute the Laplace approximation over $g$, we must obtain the mode $g^\star=\underset{g}{\text{argmax}}\ P(\bm{Y},g|m_s^c)$, the root of the equation 

\begin{equation}
-Qb^{2}(P+3)g^{3}+(b(Nb-P-4)-2Q)g^{2}+(Nb(2-R^{2})-3)g+N:=0
\end{equation}

where $Q=1-R^2$. We also require the Hessian evaluated at the mode,

\begin{align*}
H^\star &= \frac{\partial^{2}}{\partial g^{2}}[\log((1+bg)^{\frac{Nb-P-1}{2}}(1+Qbg)^{-\frac{Nb-1}{2}}g^{-\frac{3}{2}}\exp\{-\frac{N}{2g}\})] \bigg{|}^{g=g^{\star}} \\
&=\frac{1}{2}\left[\frac{(Nb-1)b^{2}Q^{2}}{(1+Qbg^\star)^{2}}-\frac{(Nb-P-1)b^{2}}{(1+bg^\star)^{2}}+\frac{3}{(g^\star)^{2}}-\frac{2N}{(g^\star)^{3}}\right].
\end{align*}

These expressions are appropriate to use in homoscedastic classes with either global or distinct regression effects; we simply compute the corresponding $R^2$ and use the appropriate $P$ based on the model under consideration. 

In classes with heteroscedasticity, the integral is intractable over both $\boldsymbol{\varphi}$ and $g$; thus we must employ a three-dimensional Laplace approximation to evaluate this integral. For computational ease and to improve the accuracy of the approximation, we again parametrize with respect to the log-variance; let $\Lambda$ represent the log-variance matrix. Denote $H_\Lambda=\Lambda\bm{1}(\bm{1}^T\Lambda\bm{1})^{-1}\bm{1}^T\Lambda$; then integrating out the global intercept and regression effects yields an expression for $P^b(\bm{Y},\lambda_{1},\lambda_{2},g|m_{s}^{c})$:

\begin{align*}
&=(2\pi)^{-\frac{Nb+P-1}{2}}\lambda_{1}^{\frac{n_{1}b}{2}-1}\lambda_{2}^{\frac{n_{2}b}{2}-1}g^{-\frac{P}{2}}b^{-\frac{P+1}{2}}|X^{T}\Lambda X|^{\frac{1}{2}}|\bm{1}^{T}\Lambda\bm{1}|^{-\frac{1}{2}}|\left(\frac{1+bg}{bg}\right)X^{T}\Lambda X-X^{T}H_{\Lambda}X|^{-\frac{1}{2}}\times \\
& J_{\Lambda}\cdot \exp\{-\frac{b}{2}[\bm{Y}\Lambda\bm{Y}-\bm{Y}^{T}H_{\Lambda}\bm{Y}-\bm{Y}^{T}(\Lambda-H_{\Lambda})^{T}X(\frac{1+bg}{bg}X^{T}\Lambda X-X^{T}H_{\Lambda}X)X^{T}(\Lambda-H_{\Lambda})\bm{Y}^{T}]\}
\end{align*}

The joint mode $(\lambda_1^\star,\,\lambda_2^\star,\,g^\star)$ is computed using the function \texttt{optim} in R; similarly, the Hessian is computed at this value using the function \texttt{hessian} in the package \texttt{numderiv}. 

\subsubsection{Heteroscedastic Zellner-Siow Cauchy Result}

Let $\boldsymbol{\beta}|\Phi,\,g\sim N(\bm{0},\,g(X^{T}\Phi X)^{-1})$
and $g\sim\text{IG}(\frac{1}{2},\,\frac{N}{2})$. Then $P(\boldsymbol{\beta}|g,\,\Phi)=(2\pi)^{-\frac{P}{2}}|g(X^{T}\Phi X)^{-1}|^{-\frac{1}{2}}\exp\{-\frac{1}{2}\boldsymbol{\beta}\frac{X^{T}\Phi X}{g}\boldsymbol{\beta}\}$
and $P(g)=\frac{\left(\frac{N}{2}\right)^{\frac{1}{2}}}{\Gamma\left(\frac{1}{2}\right)}g^{-\frac{3}{2}}\exp\{-\frac{N}{2g}\}$. 

\begin{eqnarray*}
P(\boldsymbol{\beta}|\boldsymbol{\varphi}) & = & \int P(\boldsymbol{\beta}|\Phi,\,g)P(g)dg\\
 & = & \int(2\pi)^{-\frac{P}{2}}|g(X^{T}\Phi X)^{-1}|^{-\frac{1}{2}}\exp\{-\frac{1}{2}\boldsymbol{\beta}\frac{X^{T}\Phi X}{g}\boldsymbol{\beta}\}\times\frac{\left(\frac{N}{2}\right)^{\frac{1}{2}}}{\Gamma\left(\frac{1}{2}\right)}g^{-\frac{3}{2}}\exp\{-\frac{N}{2g}\}dg\\
 & = & \int(2\pi)^{-\frac{P}{2}}|X^{T}\Phi X|^{\frac{1}{2}}(\frac{N}{2})^{\frac{1}{2}}\pi^{-\frac{1}{2}}g^{-\frac{P+3}{2}}\exp\{-\frac{1}{2}\boldsymbol{\beta}^{T}\frac{X^{T}\Phi X}{g}\boldsymbol{\beta}-\frac{1}{2}\cdot\frac{N}{g}\}dg\\
 & = & \int(2\pi)^{-\frac{P}{2}}|X^{T}\Phi X|^{\frac{1}{2}}N^{\frac{1}{2}}2^{-\frac{1}{2}}\pi^{-\frac{1}{2}}\times g^{-\frac{P+1}{2}-1}\exp\{-\frac{\nicefrac{\boldsymbol{\beta}^{T}X^{T}\Phi X\boldsymbol{\beta}+N}{2}}{g}\}dg\\
 & = & 2^{-\frac{P+1}{2}}\pi^{-\frac{P+1}{2}}|X^{T}\Phi X|^{\frac{1}{2}}N^{\frac{1}{2}}[(\boldsymbol{\beta}^{T}X^{T}\Phi X\boldsymbol{\beta}+N)^{\frac{P+1}{2}}2^{-\frac{P+1}{2}}\Gamma\left(\frac{P+1}{2}\right)]^{-1}\\
 & = & \pi^{-\frac{P+1}{2}}|X^{T}\Phi X|^{\frac{1}{2}}N^{\frac{1}{2}}(\boldsymbol{\beta}^{T}X^{T}\Phi X\boldsymbol{\beta}+N)^{-\frac{P+1}{2}}\Gamma\left(\frac{P+1}{2}\right){}^{-1}\\
 & = & \pi^{-\frac{P+1}{2}}|X^{T}\Phi X|^{\frac{1}{2}}N^{\frac{1}{2}}(\frac{\boldsymbol{\beta}^{T}X^{T}\Phi X\boldsymbol{\beta}+N}{N})^{-\frac{P+1}{2}}N^{-\frac{P+1}{2}}\Gamma\left(\frac{P+1}{2}\right){}^{-1}\\
 & = & \pi^{-\frac{P+1}{2}}|X^{T}\Phi X|^{\frac{1}{2}}N^{-\frac{P}{2}}(1+\boldsymbol{\beta}^{T}\frac{X^{T}\Phi X}{N}\boldsymbol{\beta})^{-\frac{P+1}{2}}\Gamma\left(\frac{P+1}{2}\right){}^{-1}\\
 & = & \pi^{-\frac{p+1}{2}}\Gamma\left(\frac{P+1}{2}\right){}^{-1}\times|\frac{X^{T}\Phi X}{N}|^{\frac{1}{2}}(1+\boldsymbol{\beta}^{T}\frac{X^{T}\Phi X}{N}\boldsymbol{\beta})^{-\frac{P+1}{2}}\\
 & \sim & \text{MVCauchy}_{P}(\text{location}=0,\,\text{scale}=\left(\frac{X^{T}\Phi X}{N}\right)^{-1}).\ \square
\end{eqnarray*}

\subsection{Supplemental ANCOVA Simulation Study}

We let $n_k:= 10$ for each level of the SLGF. The parameter settings are provided in Table \ref{table:ancovasettings}. 

\begin{figure}[htp]
\begin{centering}
\includegraphics[scale=.7]{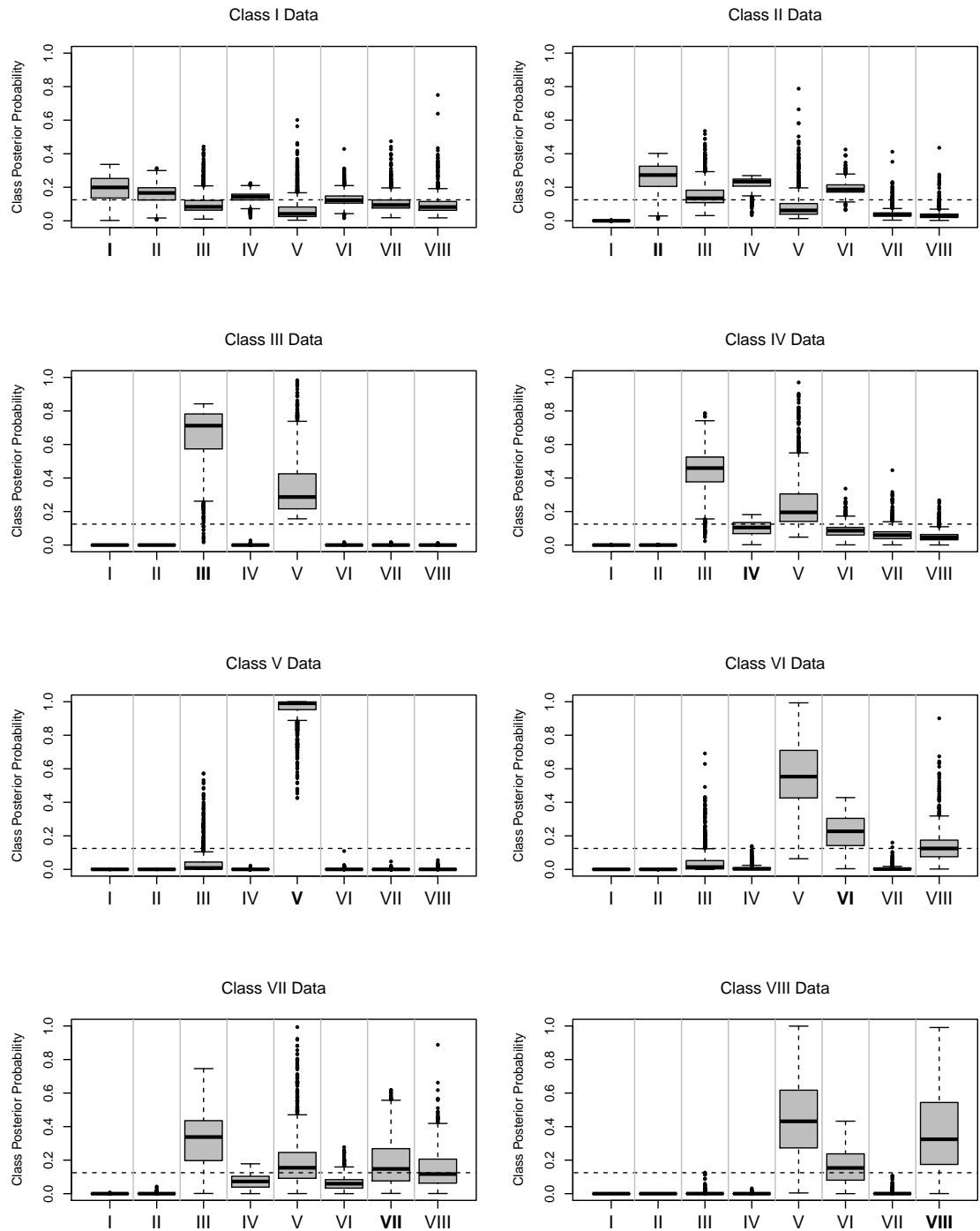}
\caption{Posterior probabilities ($y$-axis) by class based on 1000 Monte Carlo data sets with $K=4$ levels
of the categorical predictor, each with 10 observations, for a total of $N=40$ observations. The true
model class is emphasized in bold on the $x$-axis. The dashed line indicates the prior by model class.}\label{fig:ancovapostprobs}
\end{centering}
\end{figure}

\subsection{Supplemental Two-way Layout Simulation Study}

We provide three additional simulation studies in the twoway layout scenario: $10\times 5$ layouts with a smaller effect size than the study provided in Section \ref{sec:twowaysimstudy}, as well as $5\times 5$ studies with larger and smaller effect sizes. 

\begin{center}
{\small{}}
\begin{table} 
\begin{centering}
{\small{}}%
\begin{tabular}{|>{\centering}p{5cm}|c||c|}
\hline 
{\small{}Class} & \multicolumn{2}{c|}{Parameters}\tabularnewline
\hline 
\multicolumn{1}{|c|}{{\small{}I (Additive Model)}} & \multicolumn{2}{c|}{{\small{}$\boldsymbol{\nu}\in\{1,\,2,\,3,\,4,\,5,\,7,\,8,\,9,\,10\}$,
$\boldsymbol{\tau}\in\{1,\,2,\,3,\,4,\,5\}$, $\sigma^{2}=1$}}\tabularnewline
\hline 
{\small{}II (Group-by-Column } & \multicolumn{2}{c|}{{\small{}$\boldsymbol{\nu}\in\{1,\,2,\,3,\,4,\,5,\,7,\,8,\,9,\,10\}$,
$\boldsymbol{\tau}_{1}\in\{1.0,\,1.5,\,2.0,\,2.5,\,3.0\}$, }}\tabularnewline
{\small{}Interaction)} & \multicolumn{2}{c|}{{\small{}$\boldsymbol{\tau}_{2}\in\{3.0,\,2.5,\,2.0,\,1.5,\,1.0\}$,
$\sigma^{2}=1$}}\tabularnewline
\hline 
{\small{}III (Heteroscedastic } & \multicolumn{2}{c|}{{\small{}$\boldsymbol{\nu}\in\{1,\,2,\,3,\,4,\,5,\,7,\,8,\,9,\,10\}$,
$\boldsymbol{\tau}\in\{1,\,2,\,3,\,4,\,5\}$, }}\tabularnewline
{\small{}Additive)} & \multicolumn{2}{c|}{{\small{}$\sigma_{1}^{2}=1.0$, $\sigma_{2}^{2}=0.25$}}\tabularnewline
\hline 
{\small{}IV (Heteroscedastic Group-} & \multicolumn{2}{c|}{{\small{}$\boldsymbol{\nu}\in\{1,\,2,\,3,\,4,\,5,\,7,\,8,\,9,\,10\}$,
$\boldsymbol{\tau}_{1}\in\{1.0,\,1.5,\,2.0,\,2.5,\,3.0\}$,}}\tabularnewline
{\small{}by-Column Interaction)} & \multicolumn{2}{c|}{{\small{}$\boldsymbol{\tau}_{2}\in\{3.0,\,2.5,\,2.0,\,1.5,\,1.0\}$,
$\sigma_{1}^{2}=1.0$, $\sigma_{2}^{2}=0.25$}}\tabularnewline
\hline 
\end{tabular}
\par\end{centering}{\small \par}
{\small{}\caption{Settings for the four model classes in the $10\times 5$ two-way layout simulation study with smaller effect size.}\label{table:twowaysettings}
}{\small \par}
\end{table}
\par\end{center}{\small \par}

\begin{figure}[htp] 
\begin{centering}
\includegraphics[scale=.7]{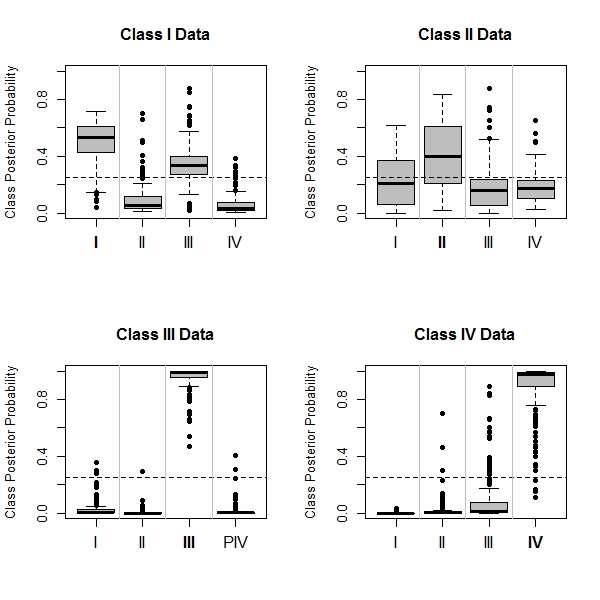}
\caption{Posterior probabilities ($y$-axis) by class based on 1000 Monte Carlo $10\times 5$ layouts with smaller effect size. The true model class is emphasized in bold on the $x$-axis. The dashed line indicates the prior by model class.}\label{fig:twoway105S}
\end{centering}
\end{figure}



\begin{center}
{\small{}}
\begin{table} 
\begin{centering}
{\small{}}%
\begin{tabular}{|>{\centering}p{5cm}|c||c|}
\hline 
{\small{}Class} & \multicolumn{2}{c|}{Parameters}\tabularnewline
\hline 
\multicolumn{1}{|c|}{{\small{}I (Additive Model)}} & \multicolumn{2}{c|}{{\small{}$\boldsymbol{\nu}\in\{1,\,1.5,\,2,\,2.5,\,3\}$, $\boldsymbol{\tau}\in\{1,\,2,\,3,\,4,\,5\}$,
$\sigma^{2}=1$}}\tabularnewline
\hline 
{\small{}II (Group-by-Column } & \multicolumn{2}{c|}{{\small{}$\boldsymbol{\nu}\in\{1,\,2,\,3,\,4,\,5\}$, $\boldsymbol{\tau}_{1}\in\{1.0,\,1.8,\,2.6,\,3.4,\,4.2\}$, }}\tabularnewline
{\small{}Interaction)} & \multicolumn{2}{c|}{{\small{}$\boldsymbol{\tau}_{2}\in\{4.2,\,3.4,\,2.6,\,1.8,\,1.0\}$,
$\sigma^{2}=1$}}\tabularnewline
\hline 
{\small{}III (Heteroscedastic } & \multicolumn{2}{c|}{{\small{}$\boldsymbol{\nu}\in\{1,\,2,\,3,\,4,\,5\}$, $\boldsymbol{\tau}\in\{1,\,2,\,3,\,4,\,5\}$, }}\tabularnewline
{\small{}Additive)} & \multicolumn{2}{c|}{{\small{}$\sigma_{1}^{2}=1.0$, $\sigma_{2}^{2}=0.25$}}\tabularnewline
\hline 
{\small{}IV (Heteroscedastic Group-} & \multicolumn{2}{c|}{{\small{}$\boldsymbol{\nu}\in\{1,\,2,\,3,\,4,\,5\}$, $\boldsymbol{\tau}_{1}\in\{1.0,\,1.8,\,2.6,\,3.4,\,4.2\}$,}}\tabularnewline
{\small{}by-Column Interaction)} & \multicolumn{2}{c|}{{\small{}$\boldsymbol{\tau}_{2}\in\{4.2,\,3.4,\,2.6,\,1.8,\,1.0\}$,
$\sigma_{1}^{2}=1.0$, $\sigma_{2}^{2}=0.25$}}\tabularnewline
\hline 
\end{tabular}
\par\end{centering}{\small \par}
{\small{}\caption{Settings for the four model classes in the two-way layout simulation study with $5\times 5$ layouts with larger effect size.}
}{\small \par}
\end{table}
\par\end{center}{\small \par}

\begin{figure}[htp] 
\begin{centering}
\includegraphics[scale=.7]{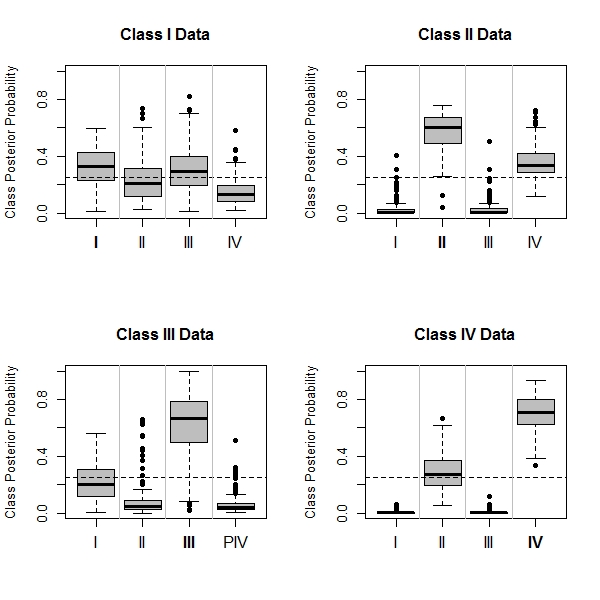}
\caption{Posterior probabilities ($y$-axis) by class based on 1000 Monte Carlo $5\times 5$ layouts with larger effect size. The true model class is emphasized in bold on the $x$-axis. The dashed line indicates the prior by model class.}\label{fig:twoway55L}
\end{centering}
\end{figure}


\begin{center}
{\small{}}
\begin{table} 
\begin{centering}
{\small{}}%
\begin{tabular}{|>{\centering}p{5cm}|c||c|}
\hline 
{\small{}Class} & \multicolumn{2}{c|}{Parameters}\tabularnewline
\hline 
\multicolumn{1}{|c|}{{\small{}I (Additive Model)}} & \multicolumn{2}{c|}{{\small{}$\boldsymbol{\nu}\in\{1,\,2,\,3,\,4,\,5\}$, $\boldsymbol{\tau}\in\{1,\,2,\,3,\,4,\,5\}$,
$\sigma^{2}=1$}}\tabularnewline
\hline 
{\small{}II (Group-by-Column } & \multicolumn{2}{c|}{{\small{}$\boldsymbol{\nu}\in\{1,\,2,\,3,\,4,\,5\}$, $\boldsymbol{\tau}_{1}\in\{1.0,\,1.5,\,2.0,\,2.5,\,3.0\}$, }}\tabularnewline
{\small{}Interaction)} & \multicolumn{2}{c|}{{\small{}$\boldsymbol{\tau}_{2}\in\{3.0,\,2.5,\,2.0,\,1.5,\,1.0\}$,
$\sigma^{2}=1$}}\tabularnewline
\hline 
{\small{}III (Heteroscedastic } & \multicolumn{2}{c|}{{\small{}$\boldsymbol{\nu}\in\{1,\,2,\,3,\,4,\,5\}$, $\boldsymbol{\tau}\in\{1,\,2,\,3,\,4,\,5\}$, }}\tabularnewline
{\small{}Additive)} & \multicolumn{2}{c|}{{\small{}$\sigma_{1}^{2}=1.0$, $\sigma_{2}^{2}=0.25$}}\tabularnewline
\hline 
{\small{}IV (Heteroscedastic Group-} & \multicolumn{2}{c|}{{\small{}$\boldsymbol{\nu}\in\{1,\,2,\,3,\,4,\,5\}$, $\boldsymbol{\tau}_{1}\in\{1.0,\,1.5,\,2.0,\,2.5,\,3.0\}$,}}\tabularnewline
{\small{}by-Column Interaction)} & \multicolumn{2}{c|}{{\small{}$\boldsymbol{\tau}_{2}\in\{3.0,\,2.5,\,2.0,\,1.5,\,1.0\}$,
$\sigma_{1}^{2}=1.0$, $\sigma_{2}^{2}=0.25$}}\tabularnewline
\hline 
\end{tabular}
\par\end{centering}{\small \par}
{\small{}\caption{Settings for the four model classes in the two-way layout simulation study with $5\times 5$ layouts with smaller effect size.}
}{\small \par}
\end{table}
\par\end{center}{\small \par}

\begin{figure}[t] 
\begin{centering}
\includegraphics[scale=.8]{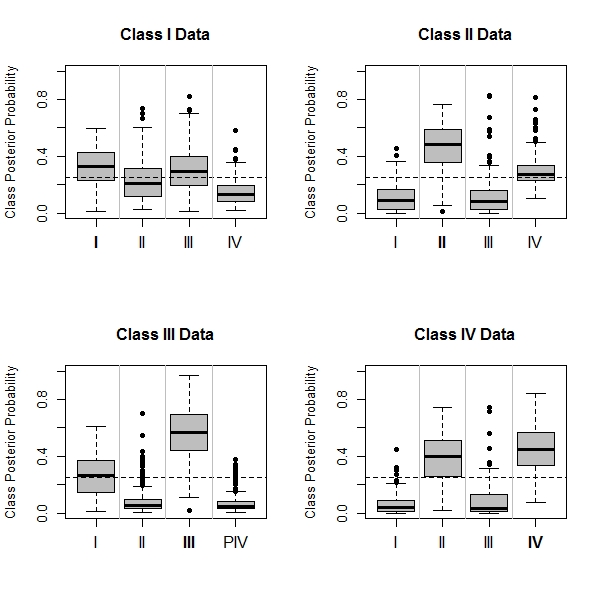}
\caption{Posterior probabilities ($y$-axis) by class based on 1000 Monte Carlo $5\times 5$ layouts with smaller effect size. The true model class is emphasized in bold on the $x$-axis. The dashed line indicates the prior by model class.}\label{fig:twoway55S}
\end{centering}
\end{figure}

\bibliographystyle{apalike}
\bibliography{TechnometricsBibliography}

\end{document}